\documentclass[draftcls,onecolumn]{IEEEtran} \def\onetwocol#1#2{#1}

\usepackage{cite}      % Written by Donald Arseneau

\usepackage{subfigure}
\usepackage{amsmath,mathrsfs} 
\usepackage{amssymb} 
\usepackage{amsfonts}
\usepackage{mathtools}

\usepackage{epstopdf}
\usepackage{ifthen}
\usepackage{amsmath}   % From the American Mathematical Society
\newcommand{\inZ}{\in{\mathbb Z}}
\newcommand{\inR}{\in{\mathbb R}}

\newcommand{\C}{ \mathbb{C}}
\newcommand{\R}{ \mathbb{R}}
\newcommand{\Z}{ \mathbb{Z}}
\newcommand{\N}{ \mathbb{N}}

\newcommand{\Lop}{{\rm L}}

\newcommand{\Top}{{\rm T}}
\newcommand{\Dop}{{\rm D}}
\newcommand{\Iop}{{\rm I}}
\newcommand{\Identity}{{\rm Id}}
\newcommand{\dint}{{\rm d}}
\newcommand{\Mop}{{\rm M}}

\newcommand{\Fourier}{ \mathcal{F}}

\newcommand{\bw}{{\boldsymbol \omega}}
\newcommand{\s}{{\zeta}}

\newtheorem{definition}{Definition}
\newtheorem{proposition}{Proposition}

\newtheorem{property}{Property}

\newtheorem{theorem}{Theorem}

\usepackage{color}
\usepackage{changebar}
\long\def\revision#1{ #1}

\def\Spc#1{{\mathcal{#1}}}  % spaces

\def\One{\mathbf{1}}    % indicator function
\def\RR{\mathbb{R}}     % Real numbers
\def\Exp{\mathbb{E}}

\def\dint{\mathrm{d}}

\def\Form{ \widehat {\mathscr{P}}}        % Characteristic form
\def\Meas{  {\mathscr{P}}}        % Characteristic form
\def\E{\mathbb E}        % Expectation
\def\Corr{\mathcal B}        % Correlation form

\begin{document}

%\title{Sparsity, fractals and generalized Poisson processes}
\title{A unified formulation of Gaussian vs. \onetwocol{\\}{\\} sparse stochastic processes---\\
Part I: Continuous-domain theory
%\\Part II: discrete-domain theory
}
%\title{Towards a unification of continuous-time\onetwocol{}{\\} and discrete theories of Gaussian and non-Gaussian stochastic processes \}

\author{Michael~Unser,~\IEEEmembership{Fellow,~IEEE,} Pouya Tafti,~\IEEEmembership{Member,~IEEE,} and Qiyu Sun
\thanks{The first and second authors are with the Biomedical Imaging Group (BIG), \'Ecole Polytechnique F\'ed\'erale de Lausanne (EPFL), CH-1015 Lausanne, Switzerland. Q. Sun is with the Department of Mathematics, University of Central Florida, Orlando, FL 32816, USA.}
}
% note the % following the last \IEEEmembership and also the first \thanks - 
% these prevent an unwanted space from occurring between the last author name
% and the end of the author line. 

\onetwocol{ }{\markboth{IEEE Transactions on Signal Processing,~Vol.~X, No.~XX,~2012}{%Shell \MakeLowercase{\textit{et al.}}: Bare Demo of IEEEtran.cls for Journals
}}
% The only time the second header will appear is for the odd numbered pages
% after the title page when using the twoside option.
% 
% *** Note that you probably will NOT want to include the author's name in ***
% *** the headers of peer review papers.                                   ***

% If you want to put a publisher's ID mark on the page
% (can leave text blank if you just want to see how the
% text height on the first page will be reduced by IEEE)
%\pubid{0000--0000/00\$00.00~\copyright~2005 IEEE}

% use only for invited papers
%\specialpapernotice{(Invited Paper)}

% make the title area
\maketitle

\begin{abstract}
We introduce a general distributional framework that results in a unifying description and characterization of a rich variety of continuous-time stochastic processes. The cornerstone of our approach is an innovation model that is driven by some generalized white noise process, which may be Gaussian or not (e.g., Laplace, impulsive Poisson or alpha stable).
This allows for a conceptual decoupling between the correlation properties of the process, which are imposed by the whitening operator $\Lop$, and its sparsity pattern which is determined by the type of noise excitation. The latter is fully specified by a L\'evy measure. %Based on the L\'evy-Khinchine formula, 
We show that the range of admissible innovation behavior varies between the purely Gaussian and super-sparse extremes.
We prove that the corresponding generalized stochastic processes are well-defined mathematically provided that the (adjoint) inverse of the whitening operator satisfies some $L_p$ bound for $p\ge1$. We present a novel operator-based method that yields an explicit characterization of all L\'evy-driven processes that are solutions of constant-coefficient stochastic differential equations (SDE). When the underlying system is stable, we recover the family of stationary CARMA processes, including the Gaussian ones. The approach remains valid when the system is unstable and leads to the identification of potentially useful generalizations of the L\'evy processes, which are sparse and non-stationary. 
Finally, we show that these processes admit a sparse representation in some matched wavelet domain and provide a full characterization of their transform-domain statistics. 
%Thier wavelet coefficients follow infinitely-divisible distributions which are heavier tailed than a Gaussian, unless the process is Gaussian.
%
%
%Finally, we show how we can apply finite difference operators to obtain a stationary characterization of these processes that is maximally decoupled and stable, irrespective of the location of the poles in the complex plane.
%to obtain a stationary characterization of these processes that is maximally decoupled and stable, irrespective of the location of the poles in the complex plane.
%The conceptual advantage is that is allows us to draw a bridge between the classical theory of Gaussian stationary processes and the more advanced theory of the L\'evy processes and to 
\end{abstract}
%\begin{keywords}
% complex-valued wavelets, Laplacian-of-Gaussian
%\end{keywords}

\section{Introduction}

%Understanding the issues of pseudo-deterministic trends (polynomials) and sinusoids that may be added to the process as the result of a non-empty but finite dimensional null space of $\Lop$.
%I suspect that there is a connection with the Wold's decomposition.
%
In recent years, the research focus in signal processing has shifted away from the classical linear paradigm, which is intimately linked with the theory of stationary Gaussian processes \cite{Papoulis1991,Gray2004}. Instead of considering Fourier transforms and performing quadratic optimization, researchers are presently favoring wavelet-like representations and have adopted ÓsparsityÓ as design paradigm \cite{Candes2008,Bruckstein2009,Mallat2009,Starck2010,Elad2010b}. The property that a signal admits a sparse expansion can be exploited elegantly for compressive sensing, which is presently a very active area of research (cf. special issue of the Proceedings of the IEEE \cite{Baraniuk2010,Elad2010}).
The concept is equally helpful for solving  inverse problems and has resulted in significant algorithmic advances for the efficient resolution of large scale $\ell_1$-norm minimization problems \cite{Figueiredo2003,Daubechies2004,Beck2009b}.
%As remarkable as these recent developments are, it remains that t

The current formulations of compressed sensing and sparse signal recovery are fundamentally deterministic. 
By drawing on the analogy with the classical theory of signal processing, it is likely that further progress may be achieved by adopting a statistical (or estimation theoretic) point of view.
This stands as our primary motivation for the investigation of the present class of continuous-time stochastic processes, the greater part of which is sparse by construction. These processes are specified as a superset of the Gaussian ones, which is essential for maintaining backward compatibility with traditional statistical signal processing.

%The cornerstone for such an approach could be the extended class of continuous-time stochastic models put forth in this series of papers. The main conceptual advantage is that these processes can display a variety of sparsity patterns, while maintaining  backward compatibility with the traditional (stationary Gaussian) models because they are all ruled by the same type of stochastic differential equations (SDE).

The inspiration for this work is provided by the innovation approach to system modeling---a standard technique in statistics and control theory that is well developed in the discrete setting and often favored by engineers. 
Innovation models are also used in signal processing for the investigation of continuous-time stationary Gaussian stochastic processes \cite{Kailath1970,Papoulis1991}.
Non-Gaussian variants of such models are easy to set up in the discrete world, but they do result in harder identification problems \cite{Giannakis1990,Swami1990,Rao1992}.
%An active topic of research is the determination of a proper noise input to simulate signals with prescribed marginal distributions  \cite{Pawlak2010, Picinbono2010}.
%
%While this kind of generative model is very simple, 
%Non-gaussian extensions of such models have also been considered
%There is no major difficult in extending those models for non-Gaussian processes
% except of course, that they become much harder to identify and to properly tune 
%
%Extending the approach for non-Gaussian processes
%In the discrete domain, it is relatively easy processes by simply switching to a non-Gaussian excitation (any type of i.i.d. random variable).
%Needless to say that the task of reversing such models is much more challenge than in the Gaussian case %provided by any type of  
%There is also an extension of the discrete theory of ARMA processes that relies on
%non-Gaussian innovations; that is, i.i.d. random variables that are fed into a LSI system. 
By contrast, there is comparatively little work on continuous-domain
innovations for the specification of non-Gaussian or/and non-stationary
processes due to the inherent difficulty of rigorously defining non-Gaussian
white noise in the continuous domain. The proper mathematical framework exists
and was developed by the Russian school of mathematics in the
1960s\cite{Gelfand-Villenkin1964},
%\revision{and also by the French and Japanese schools \cite{Schwartz1973,Ito1984}}
but has hardly been used by
practitioners until now. This is mainly due to the widespread acceptance of
stochastic integration (It\^o calculus) in the advanced theory of stochastic
processes  \cite{Karatzas1991,Okensal2007,Samorodnitsky1994, Appelbaum2009},
which avoids the direct handling of white noise and tempered distributions.

By following up on our initial work on the generation of piecewise-smooth signals from random streams of Dirac impulses (Poisson white noise)  \cite{Unser2011}, our present aim is to set the foundations of a comprehensive theory of continuous-domain stochastic processes based on the simple, unifying principle of the filtering of special brands of (non-Gaussian) white noise. While the concept remains applicable in multiple dimensions, we focus on the time domain (1-D signals),  and provide a systematic treatment of systems that are described by ordinary differential equations, including some novel twists for the non-stable scenarios, which opens the door to interesting generalizations.
%, the latter leading to the most relevant configurations for a wavelet-based analysis.
The primary contributions are:
\begin{enumerate}
%{  \rule{0.5ex}{0.5ex}}	{\topsep=0.5ex \leftmargin=1.3em \itemsep=-0.4ex \rightmargin=0cm}
\item The extension of our prior innovation models to the broadest possible class of white noises beyond the Gaussian and impulsive Poisson categories: We show that each brand is uniquely specified by a
 L\'evy measure that conditions the degree of sparsity of the process. The Gaussian processes are the least sparse ones; the Poisson processes are intermediate with their level of sparsity being controlled by the rate parameter $\lambda$\cite{Unser2011}. The sparsest processes are the alpha-stable ones whose marginal distributions are heavy tailed with unbounded variance \cite{Shao1993,Samorodnitsky1994}. 
% we also provide a solid argumentation as to why the non-Gaussian brands noise are intrinsically sparse with their level of sparsity is directly determined by the tail behavior of the L\'evy density.
 \item The systematic investigation of processes that are ruled by constant-coefficient SDEs together with the proposal of a generic operator-based method of solution: When the underlying system is stable, we recover the complete family of (non-Gaussian) continuous-time autoregressive moving average (CARMA) processes (see also the work of Brockwell for an equivalent state-space characterization that relies on stochastic integrals \cite{Brockwell2001}). The further reaching aspect of our formulation is that the method remains applicable in the non-stable case and that it leads to some interesting generalizations of L\'evy processes, which are non-stationary. 
 %Note that the classical L\'evy  processes are recovered as one of the simplest instances of our theory (single pole at the origin).
 
 \item The generalization/extension of our previous stability and existence results (cf. \cite[Theorem 2]{Unser2011}, \cite[Theorem 1.3]{Sun2012}) for the present enlarged class of stochastic processes: In essence, we are replacing the basic $L_2$-boundedness requirement that is central to the continuous-time Gaussian theory by a more robust $L_p$ condition  (cf. Theorem 3); the case $p=1$ is required for the non-symmetric Poisson processes,  while the range of values $p\in(0,2)$ becomes appropriate for the alpha-stable processes. 
 %The general trend is that sparser processes require smaller values of $p$.
 \item The demonstration that these processes admit a sparse representation in some matched wavelet-like basis together with a complete characterization of the transform-domain statistics. In particular, we prove that the wavelet coefficients follow {\em infinitely divisible} probability laws that are heavier tailed than a Gaussian (whenever the innovation is non-Gaussian).
\end{enumerate}

The paper is organized as follows. The basic flavor of the innovation model is conveyed in Section II by focusing on a first-order differential system which results in the generation of Gaussian and non-Gaussian AR(1) stochastic processes.
We use of this model to illustrate that a properly-matched wavelet transform can outperform the classical Karhunen-Lo\`eve transform (or the DCT) for the compression of (non-Gaussian) signals.
In Section III, we review the foundations of Gelfand's theory of generalized stochastic processes.
% together with some basic tools and concepts in functional analysis. 
In particular, we characterize the complete class of admissible continuous-time white noise processes and give some argumentation as to why the non-Gaussian brands are inherently sparse.  In Section IV, we give a high-level description of the general innovation model and provide a novel operator-based method for the solution of SDE.
In Section V, we make use of Gelfand's formalism to fully characterize our extended class of (non-Gaussian) stochastic processes including the special cases of CARMA and $N$th-order generalized L\'evy processes. 
We also derive the statistics of the wavelet-domain representation of these signals, which allows for a common (stationary) treatment of the two latter classes of processes,  irrespective of any stability consideration.
Finally, in Section VI, we turn back to our introductory example by moving into the unstable regime (single pole at the origin) which yields a non-conventional system-theoretic interpretation of classical L\'evy processes\cite{Levy1954,Sato1994,Appelbaum2009}. We also point out the structural similarity between
the increments of L\'evy processes and their Haar wavelet coefficients. 
For higher-order illustrations of sparse processes, we refer to our companion paper \cite{Unser2012b}, which is specifically devoted to the study of the discrete-time implication of the theory and the way to best decouple (e.g.\ ``sparsify") such processes.  The notation, which is common to both papers, is summarized in \cite[Table II]{Unser2012b}.

\section{Motivation: Gaussian vs. non-Gaussian AR(1) processes}
\label{Sec:motiv}
A continuous-time Gaussian AR(1) (or Gauss-Markov) process can be formally generated by applying a first-order analog filter to a Gaussian white noise process $w$:
\begin{align}
\label{eq:AR1}
s_\alpha(t)=(\rho_\alpha \ast w)(t)
\end{align}
where $\rho_\alpha(t)=\One_+(t) e^{\alpha t}$ with ${\rm Re}(\alpha)<0$ and $\One_{+}(t)$ is the unit-step %(or Heaviside)
function. Since 
$\rho_\alpha=(\Dop-\alpha \Identity)^{-1}\delta$ where $\Dop=\frac{\dint}{\dint t}$ and $\Identity$ are the derivative and identity operators respectively, $s_\alpha$ satisfies the ``innovation" model (cf. \cite{Kailath1970,Papoulis1991})
\begin{align}
\label{eq:AR1w}
(\Dop-\alpha \Identity)s_\alpha(t)=w(t),
\end{align}
or, equivalently, the stochastic differential equation (SDE) (cf. \cite{Okensal2007})
\begin{align*}
\dint s_\alpha(t)-\alpha s_\alpha(t)\dint t=\dint W(t),
\end{align*}
where $W(t)=\int_0^t w(\tau) \dint \tau$ is a standard Brownian motion (or Wiener process) excitation. In the statistical literature, the solution of the above first-order SDE is often called the
Ornstein-Uhlenbeck process.

Let $(s_\alpha[k]=\left.s_\alpha(t)\right|_{k=t})_{k \inZ}$ denote the sampled version of the continuous-time process. Then, one can show that $s_\alpha[\cdot]$ is a discrete AR(1) autoregressive process that can be whitened by applying the first-order linear predictor:
\begin{align}
\label{Eq:predict}
s_\alpha[k]-e^{\alpha}s_\alpha[k-1]%=\Delta_\alpha s_\alpha(k)
=u[k]
\end{align}
where $u[\cdot]$ (prediction error) is an i.i.d. Gaussian sequence. Alternatively, one can decorrelate the signal by computing its discrete cosine transform (DCT), which is known to be asymptotically equivalent to the Karhunen-Lo\`eve transform (KLT) of the process \cite{Ahmed1974,Unser1984a}. Eq. \eqref{Eq:predict} provides the basis for classical linear predictive coding (LPC), while the decorrelation property of the DCT is often invoked to justify the popular JPEG transform-domain coding scheme \cite{Jayant1984}.
\begin{figure}
\centering
\includegraphics[width=8cm]{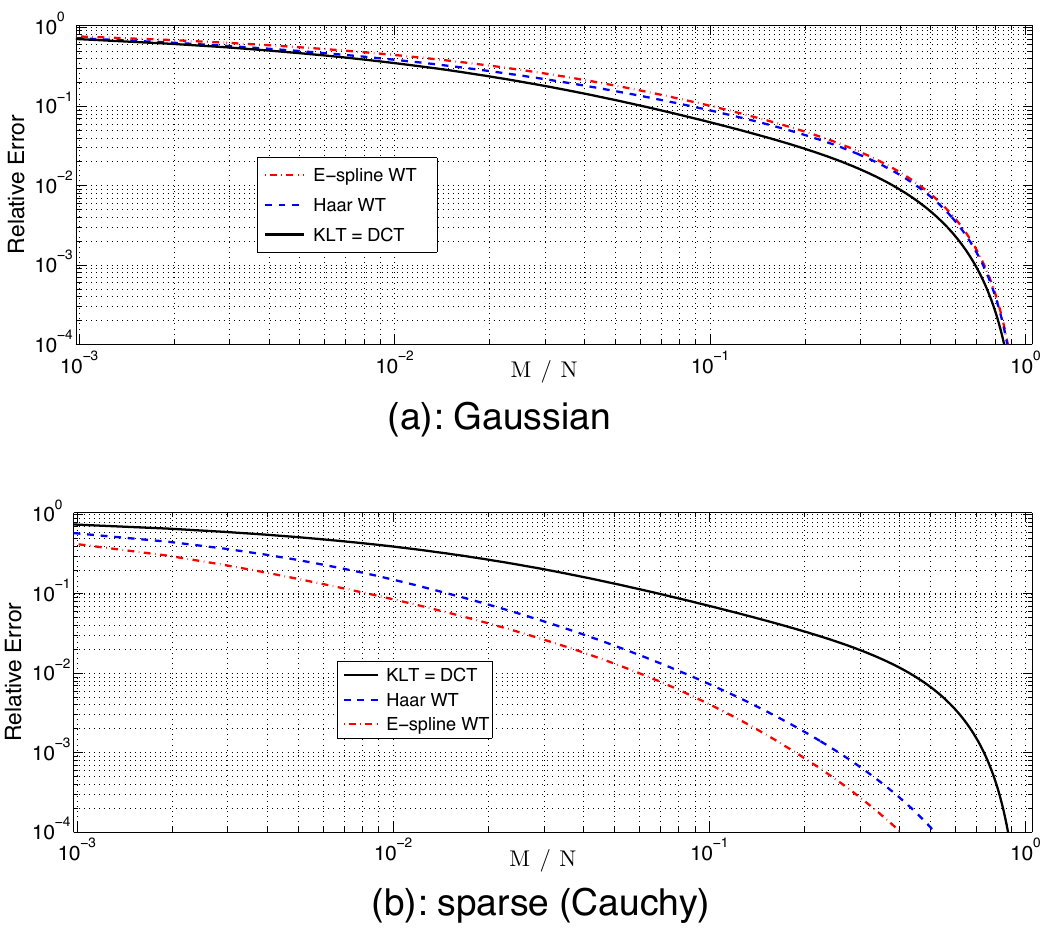}
  \caption{\label{fig:approx}
Wavelets vs. KLT (or DCT) for the $M$-term approximation of Gaussian vs.\ sparse AR(1) processes with $\alpha=-0.1$: (a) classical Gaussian scenario, (b) sparse scenario with symmetric Cauchy innovations. The E-spline wavelets are matched to the innovation model. The displayed results (relative quadratic error as a function of $M/N$) are averages over 1000 realizations for AR(1) signals of length $N=1024$; the performance of DCT and KLT is undistinguishable. }

\end{figure}

\begin{figure}
\centering
\includegraphics[width=8cm]{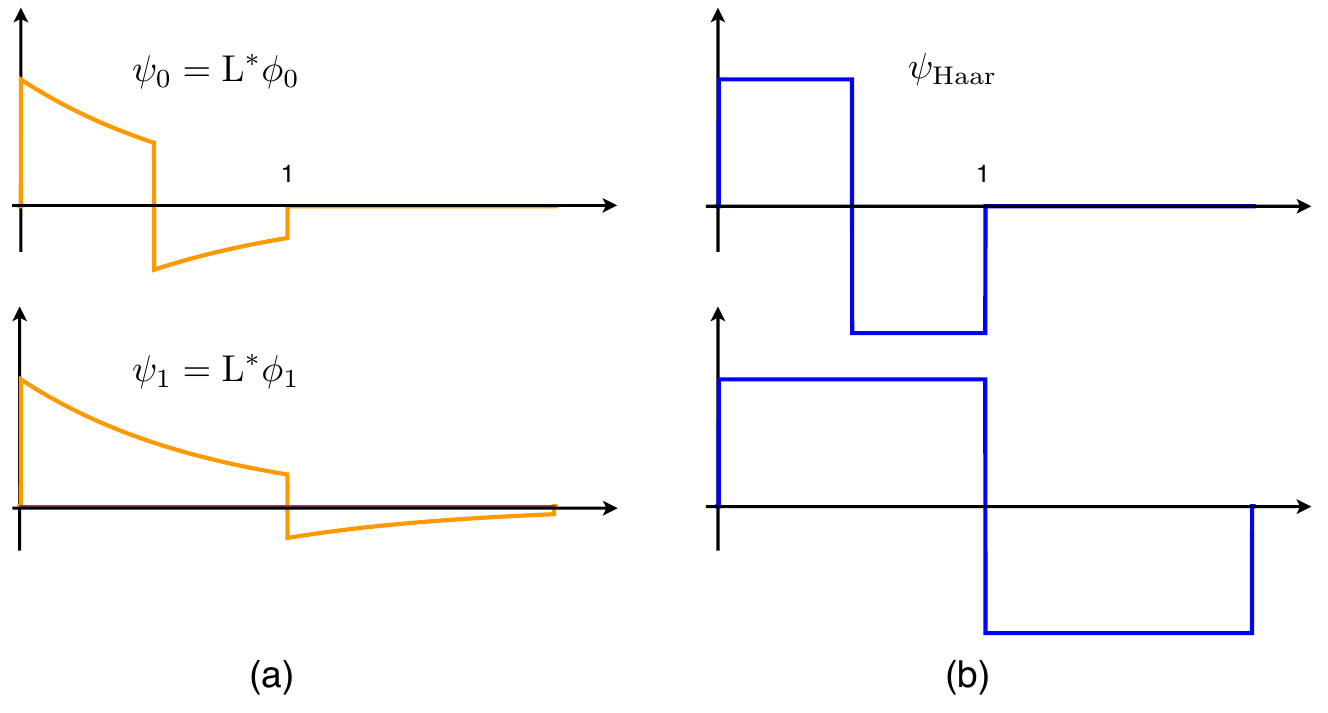}
  \caption{\label{fig:wav}
Comparison of operator-like and conventional wavelet basis functions at two successive scales: (a) first-order E-spline wavelets with $\alpha=-0.5$. (b) Haar wavelets. The vertical axis is rescaled for full range display. }

\end{figure}

In this paper, we are concerned with the non-Gaussian counterpart of this story, which, as we shall see, will result in the identification of sparse processes. The idea is to retain the simplicity of the classical innovation model, while substituting the continuous-time Gaussian noise by some generalized L\'evy innovation (to be properly defined in the sequel). This translates into Eqs. \eqref{eq:AR1}-\eqref{Eq:predict} remaining valid, except that the underlying random variates are no longer Gaussian.
%, but {\em infinitely divisible}. 
The more significant finding
%, which came as a pleasing surprise to us, 
is that the KLT (or its discrete approximation by the DCT) is no longer optimal for producing the best $M$-term approximation of the signal. This is illustrated in Fig. \ref{fig:approx}, which compares the performance of various transforms for the compression of two kinds of AR(1) processes with correlation $e^{-0.1}\approx0.90$: Gaussian vs.\ sparse where the latter innovation follows a Cauchy distribution. The key observation is that the E-spline wavelet transform, which is matched to the operator $\Lop=\Dop-\alpha \Identity$, provides the best results in the non-Gaussian scenario over the whole range of experimentation [cf. Fig.  \ref{fig:approx}(b)], while the outcome in the Gaussian case is as predicted by the classical theory with the KLT being superior. Examples of orthogonal E-spline wavelets at two successive scales are shown in Fig. \ref{fig:wav} next to their Haar counterparts. We selected the E-spline wavelets because of their ability to decouple the process which follows from their operator-like behavior: $\psi_i=\Lop^\ast \phi_i$ where $i$ is the scale index and $\phi_i$ a suitable smoothing kernel  \cite[Theorem 2]{Khalidov2006}. Unlike their conventional cousins, they are not dilated versions of each other, but rather extrapolations in the sense that the slope of the exponential segments remains the same at all scales. They can, however, be computed efficiently using a perfect reconstruction filterbank with 
scale-dependent filters \cite{Khalidov2006}.

The equivalence  with traditional wavelet analysis (Haar) and finite-differencing (as used in the computation of total variation) for signal ``sparsification''  is achieved by letting $\alpha \rightarrow 0$. The catch, however, is that the underlying system becomes unstable! Fortunately, the problem can be fixed, but it calls for an advanced mathematical treatment that is beyond the traditional formulation of stationary processes. 
The reminder of the paper is devoted to giving a proper sense to what has just been described informally, and to extending the approach to the whole class of ordinary differential operators, including the non-stable scenarios. 
The non-trivial outcome, as we shall see, is that many non-stable systems are linked with non-stationary stochastic processes. These, in turn, can be stationarized and ``sparsified" by application of a suitable wavelet transformation.
The companion paper \cite{Unser2012b} is focused on the discrete aspects of the theory including the generalization of \eqref{Eq:predict} for decoupling purposes and the full characterization of the underlying processes.
\section{Mathematical background}

The purpose of this section is to introduce the distributional formalism that is required for the proper definition of continuous-time white noise that is the driving term of \eqref{eq:AR1} and its generalization.
%that was developed by the Russian school of mathematics in the 1960s for the study of continuous-time stochastic processes and to properly define the concept of white noise. 
We start with a brief summary of some required notions in functional analysis, which also serves us to set the notation.
We then introduce the fundamental concept of characteristic functional which constitutes the foundation of Gelfand's theory of generalized stochastic processes. We proceed by giving the complete characterization of the possible types of continuous-domain white noises---not necessarily Gaussian---which will be used as universal input for our innovation models. We conclude the section by showing that the non-Gaussian brands of noises that are allowed by Gelfand's formulation are intrinsically sparse, a property that has not been emphasized before (to the best of our knowledge).

\subsection{Functional and distributional context}

%Our signals are real-valued
%functions on the real line with complex-valued Fourier transforms.
%%, although all definitions are easily extensible to functions on $\RR^d$.  Note to
%% mathematicians: as is common in engineering and also to some extent in
%% physics, we shall frequently 
%We shall denote a function $f:\RR\to\CC$ or a generalized
%function on $\RR$ as $f(t)$ and its Fourier transform as $\hat{f}(\omega)$.
%In this usage, $t$ and $\omega$ indicate the domain, not a point
%at which the function is evaluated.}

%The basic Lebesgue $L_p$ spaces are denoted by
%$$L_p=\left\{ f(t): \|f\|_p < +\infty\right\}$$
The $L_p$-norm of \revision{a function $f=f(t)$ is }
 $
 \|f\|_p=\left(\int_\R |f(t)|^p \dint t\right)^\frac{1}{p}$ for $1 \le p < \infty$ and
$\|f\|_\infty={\rm ess} \sup_{t \inR} |f(t)|$ for $p=+\infty$  with the
corresponding Lebesgue space being denoted by \revision{$L_p=L_p(\RR)$}. The concept is extendable for characterizing the
rate of decay of functions. To that end, we introduce the weighted $L_{p,\alpha}$ spaces with $\alpha\inR^+$
%We will characterize the decay of functions by their inclusion in the weighted $L_{p,\alpha}$ spaces with $\alpha\inR^+$:
 $$L_{p,\alpha}=\left\{ f \in L_p: \|f\|_{p,\alpha} < +\infty\right\}$$
where the $\alpha$-weighted $L_p$-norm of $f$ is defined as
\revision{\[ \|f\|_{p,\alpha}= \|(1+|\cdot|^\alpha)f(\cdot)\|_p\text{.}\]}
%=\left(\int_\R \left|(1+|t|^\alpha)f(t)\right|^p \dint t\right)^\frac{1}{p}$$ for $1 \le p < \infty$ and
%$$\|f\|_{\infty,\alpha}=\sup_{t \inR} \left\{(1+|t|^\alpha) |f(t)|\right\}$$ for $p=+\infty$.
%The traditional Lebesgue $L_p$ spaces are recovered with $\alpha=0$.
Hence, the statement $f \in L_{\infty,\alpha}$ implies that $f(t)$ \revision{decays at
least as fast as $1/|t|^\alpha$} as $t$ tends to $\pm\infty$; more precisely, that
$|f(t)|\le \frac{\|f\|_{\infty,\alpha}}{1+|t|^\alpha}$ almost everywhere. In
particular, this allows us to infer that $L_{\infty,\frac{1}{p}+\epsilon}
\subset L_p$ for any $\epsilon>0$ and $p\ge1$. Another obvious inclusion is
$L_{p,\alpha} \subseteq L_{p,\alpha_0}$ for any $\alpha \ge \alpha_0$. In the
limit, we end up with the space of rapidly-decreasing functions
\revision{$\mathcal{R}=\left\{ f:\|f\|_{\infty,m}<+\infty,\  \forall m
\inZ^+\right\}$}, which is included in all the others.%
\footnote{The topology of $\Spc R$ is defined by the family of
semi-norms $\|\cdot\|_{\infty,m}$, $m=1,2,3,\ldots$.}
%Other embedding relations are $L_\infty(\R,w_{\frac{1}{p}+\epsilon}) \subset L_p$

We use \revision{$\varphi=\varphi(t)$} to denote a generic function in Schwartz's class $\mathcal{S}$ of rapidly-decaying and infinitely-differentiable test functions. Specifically, Schwartz's space is defined as:$$\mathcal{S}=\left\{ \varphi\in C^\infty:\|\Dop^n\varphi\|_{\infty,m}<+\infty,\  \forall m,n \inZ^+\right\},$$
with the operator notation $\Dop^n=\frac{\dint^n}{\dint t^n}$ and the convention that $\Dop^0=\Identity$ (identity).
%This allows us to define Schwartz's class of infinitely-differentiable and rapidly-decreasing test functions as
$\mathcal{S}$ is a complete topological vector space. Its topological dual is the space of tempered distributions $\mathcal{S}'$; a distribution $\phi \in \mathcal{S}'$ is a continuous linear functional on $\mathcal{S}$ that is characterized by
a duality product rule $\phi(\varphi)=\langle \phi, \varphi \rangle=\int_\R \phi(t) \varphi(t) \dint t$ with $\varphi \in \mathcal{S}$ where the right-hand side expression has a literal interpretation as an integral only when $\phi(t)$ is true function of $t$. The prototypical example of a tempered distribution is the Dirac distribution $\delta$, which is defined as $\delta(\varphi)=\langle \delta, \varphi \rangle=\varphi(0)$. In the sequel, we will drop the explicit dependence of the distribution on the generic test function $\varphi\in \mathcal{S}$ and simply write $\phi$ or even $\phi(t)$ (with an abuse of notation).

Let $\Top$ be a continuous\footnote{An operator $\Top$ is continuous 
from a \revision{(sequential)} topological vector space $\mathcal{V}$ into another one iff.
$\varphi_k \rightarrow \varphi$ in the topology of $\mathcal{V}$ implies that
$\Top\varphi_k \rightarrow \Top\varphi$ in the topology (or norm) of the
second space. If the two spaces coincide, we say that $\Top$ is
$\mathcal{V}$-continuous.
%This is typically established by the operator be bounded in some appropriate norm: $\|\Top \varphi\|\le C \|\varphi\|$ for all $\varphi \in \mathcal{V}$.
}
linear operator that maps $\mathcal{S}$ into itself (or eventually some enlarged topological space such as $L_p$).
It is then possible to extend the action of $\Top$ over $\mathcal{S}'$ (or an appropriate subset of it) based on the definition
$\langle \Top \phi, \varphi \rangle=\langle \phi, \Top^\ast  \varphi \rangle$ if $\Top^\ast$ is the adjoint of $\Top$ which maps $\varphi$ to another test function $ \Top^\ast  \varphi \in \mathcal{S}$ continuously. An important example is the Fourier transform whose classical definition is 
$\Fourier\{f\}(\omega)=\hat f(\omega)=\int_{\R} f(t) e^{-j \omega t} \dint t$.
Since $\Fourier$ is a self-adjoint $\mathcal{S}$-continuous operator, it is extendable to $\mathcal{S}'$ based on the adjoint relation
$\langle \Fourier \phi, \varphi \rangle=\langle \phi, \Fourier \varphi \rangle$ for all $\varphi \in \mathcal{S}$ (generalized Fourier transform).

%We recall that $\Dop^n$ is a continuous linear operator from $\mathcal{S}$ into itself, which is shift-invariant, commutative, associative and distributive, and whose adjoint is $\Dop^{n\ast}=(-1)^n \Dop^{n}$.
%%The same also holds true for the distributional extension of the operator (weak derivatives), which defines a mapping from $\mathcal{S'}$ into itself. 
%The so-called $n$th-order {\em weak} derivative of the tempered distribution $\phi$ is defined as
%$ \Dop^{n} \phi(\varphi)=\langle \Dop^{n} \phi, \varphi \rangle=\langle \phi, \Dop^{n\ast} \varphi \rangle$ for any $\varphi \in \mathcal{S}$. The latter operator---or, by extension, any polynomial of distributional derivatives of the form $P_N(\Dop)=\sum_{n=1}^N a_n\Dop^n$ with constant coefficients $a_n \in \C$---maps $\mathcal{S'}$ into itself and enjoys the same properties as its classical counterpart: shift-invariance, commutativity, associativity and distributivity.

A linear, shift-invariant (LSI) operator that is well-defined over $\mathcal{S}$ can always be written as a convolution product:
\begin{eqnarray*}
\label{eq:conv}
\Top_{\rm LSI}\varphi(t)=(h \ast \varphi)(t)=\int_{\R} h(\tau) \varphi(t-\tau) \dint \tau
\end{eqnarray*}
where $h(t)=\Top_{\rm LSI}\delta(t)$ is the impulse response of the system.
The adjoint operator is the convolution with the time-reversed version of $h$:
$$h^\vee(t)\equiv h(-t).$$ The better-known categories of LSI operators are the BIBO-stable (bounded input, bounded output) filters, and the ordinary differential operators. While the latter are not BIBO-stable, they do work well with test functions.
\subsubsection{$L_p$-stable LSI operators} The BIBO-stable filters correspond to the case where $h \in L_1$, or, more generally, when $h$ corresponds to a complex-valued Borel measure of bounded variation. The latter extension allows for discrete filters of the form $h_d(t)=\sum_{n \inZ} d[n]\delta(t-n)$ with $d[n]\in \ell_1$. We will refer to these filters as $L_p$-stable because they are bounded in all $L_p$-norms (by Young's inequality).
%, whose impulse response is technically not in $L_1$.
$L_p$-stable convolution operators satisfy the properties of commutativity, associativity, and distributivity with respect to addition. 

\subsubsection{$\mathcal{S}$-continuous LSI operators}
For an $L_p$-stable filter to yield a Schwartz function as output, it is necessary that its impulse response (continuous or discrete) be rapidly-decaying. In fact, the condition $h \in \mathcal{R}$ (which is much stronger than integrability) ensures that the filter is $\mathcal{S}$-continuous.
The $n$th-order derivative $\Dop^n$ and its adjoint $\Dop^{n\ast}=(-1)^n \Dop^{n}$ are in the same category. The $n$th-order {\em weak} derivative of the tempered distribution $\phi$ is defined as
$ \Dop^{n} \phi(\varphi)=\langle \Dop^{n} \phi, \varphi \rangle=\langle \phi, \Dop^{n\ast} \varphi \rangle$ for any $\varphi \in \mathcal{S}$. The latter operator---or, by extension, any polynomial of distributional derivatives $P_N(\Dop)=\sum_{n=1}^N a_n\Dop^n$ with constant coefficients $a_n \in \C$---maps $\mathcal{S'}$ into itself. The class of these differential operators enjoys the same properties as its classical counterpart: shift-invariance, commutativity, associativity and distributivity.

\subsection{Notion of generalized stochastic process}
The leading idea in distribution theory is that a generalized function $\phi$ is not defined
through its point values $\phi(t), t\inR$, but rather through its scalar products $\phi(\varphi)=\langle \phi, \varphi \rangle$ with all ``test" functions
$\varphi \in \mathcal{S}$. 
In an analogous fashion, Gelfand and Vilenkin define a generalized stochastic process $s$ via the probability law
of its scalar products with arbitrary test functions $\varphi\in \mathcal{S}$ \cite{Gelfand-Villenkin1964}, rather than by
considering the probability law of its pointwise samples $\{\dots,s(t_1), s(t_2),\dots,s(t_N),\dots\}$, as is customary in the conventional formulation.

Let $s$ be such a generalized process. We first observe that the scalar product $X_1=\langle s, \varphi_1 \rangle$ with a given test function $\varphi_1$ is a conventional (scalar) random variable that is characterized by its probability density function (pdf) $p_{X_1}(x_1)$; the latter is in one-to-one correspondence (via the Fourier transform) with the characteristic function 
$\hat p_{X_1}(\omega_1)=\E\{e^{j\omega_1 X_1} \}=\int_\R e^{j\omega_1 x_1} p_{X_1}(x_1) \dint x_1=\E\{e^{j\langle s, \omega_1\varphi_1\rangle} \}$ where $\E\{\cdot\}$ is the expectation operator.
The same applies for the 2nd-order pdf $p_{X_1,X_2}(x_1,x_2)$ associated with a pair of test functions $\varphi_1$ and $\varphi_2$ which is the inverse Fourier transform of the 2-D characteristic function $\hat p_{X_1,X_2}(\omega_1,\omega_2)=\E\{e^{j\langle s, \omega_1\varphi_1 + \omega_2 \varphi_2\rangle} \}$, and so forth if one wants to specify higher-order dependencies.
%, which is positive-definite by construction (Bochner's theorem). 

The foundation for the theory of generalized stochastic processes is that one can deduce the complete statistical information about the process from the knowledge of its characteristic form
\begin{equation}
\Form_s(\varphi)=\E\{ e^{j\langle s, \varphi \rangle} \}
\end{equation}
which is a continuous, positive-definite functional over $\mathcal{S}$ such that $\Form_s(0)=1$. Since the variable $\varphi$ in $\Form_s(\varphi)$ is completely generic, it provides the equivalent of an infinite-dimensional generalization of the characteristic function. Indeed, any finite dimensional version can be recovered by direct substitution of $\varphi=\omega_1 \varphi_1 + \cdots + \omega_N \varphi_N$ in $\Form_s(\varphi)$  where the $\varphi_n$ are fixed and where $\bw=(\omega_1, \cdots,\omega_N)$ takes the role of the $N$-dimensional Fourier variable.
In fact, Gelfand's theory rests upon the principle that specifying an admissible functional $\Form_s(\varphi)$ is equivalent to defining the underlying generalized stochastic process (Bochner-Minlos theorem).
The precise statement of this result, which relies upon the fundamental notion of positive-definiteness, is given in Appendix I.

\subsection{White noise processes}
We define \revision{a white noise $w$} as a generalized random process that is
stationary and whose measurements for non-overlapping test functions are
independent. A remarkable aspect of the theory of generalized stochastic
processes is that it is possible to deduce the complete class of such noises
based on functional considerations only \cite{Gelfand-Villenkin1964}.
\label{sec:whitenoise}  
To that end, Gelfand and Vilenkin consider
the generic class of functionals of the form
\begin{equation}
\label{eq:gennoise}
\Form_w(\varphi)=\exp\left( \int_{\R} f\big(\varphi(t)\big) \dint t\right)
\end{equation}
\revision{where $f$ is a continuous function on the real line and $\varphi$ is
a test function from some suitable space. This functional specifies an
independent noise process if $\Form_w$ is continuous and positive-definite and $\Form_w(\varphi_1+\varphi_2)=\Form_w(\varphi_1)\Form_w(\varphi_2)$ whenever $\varphi_1$ and $\varphi_2$ have non-overlapping support.
%(i.e., $\varphi_1(t) \cdot \varphi_2(t)=0$).
%, which is the functional way of indicating that the process is taking independent values at every point.
The latter property is equivalent to having $f(0)=0$ in (\ref{eq:gennoise}).
Gelfand and Vilenkin then go on to prove that the complete class of
functionals of the form (\ref{eq:gennoise}) with the required mathematical
properties (positive-definitess and factorizability) is obtained
by choosing $f$ to be a {\em L\'evy exponent,\/} as defined below.

\begin{definition}
\label{Def:Levy}
A complex-valued continuous function  $f(\omega)$ is a valid L\'evy exponent if and only if $f(0)=0$ and $g_\tau(\omega)=e^{\tau f(\omega)}$ is a positive-definite function of $\omega$ for all $\tau\inR^+$.
\end{definition}
The reader who is not familiar with the notion of positive definiteness is referred to Appendix I.

%$$
%f(u)=b_0+j b_1' u - \frac{b_2 u^2}{2}+\int_{\R\backslash\{0\}} [e^{j a u} - r(a)(1+ja u)]\, R(d a)  
%$$
%where $R$ is an other arbitrary positive measure on $\R$ such that
%$
%\int_{\R\backslash\{0\}}  \min(1,a^2)\, R(d a) < \infty
%$
%and where $r(a)$ is a function such that $r(a)-1$ has a third-order zero at $a=0$ and $\int_{|a|>0} [1 - r(a)] \, R(d a)+b_0=0$ (to ensure that $f(0)=0$).
In doing so, they actually establish a one-to-one correspondence between the characteristic form of an independent noise processes (\ref{eq:gennoise}) and the family of infinite-divisible laws whose 
characteristic function takes the form $\hat p_X(\omega)=e^{f(\omega)}=\E\{e^{j \omega X}\}$ \cite{Feller1971, Steufel2003}. 
While Definition \ref{Def:Levy} is hard to exploit directly, the good news is that there exists a complete constructive, characterization of L\'evy exponents, which is a classical result in probability theory:
%provided by the L\'evy-Khinchine formula:

\begin{theorem}[L{\'e}vy-Khintchine formula]
\label{Th:LK}
$f(\omega)$ is a valid L\'evy exponent if and only if it can be written as
\begin{equation}
\label{eq:levykintchine}
f(\omega)=j b'_1 \omega - \frac{b_2 \omega^2}{2}+\int_{\R\backslash\{0\}}
[e^{j a \omega} - 1 - j a \omega \One_{\{|a|<1\}}(a)] \, V(\dint a)  
\end{equation}
where $b_1'\in \R$ and $b_2\in R^+$ are some constants and $V$ is a L\'evy
measure, that is, a (positive) Borel measure on $\R\backslash\{0\}$ such that
\begin{equation}
\int_{\R\backslash\{0\}}  \min(1, a^2) \, V(\dint a) < \infty.
\label{eq:admissibility}
\end{equation}
\end{theorem}

The notation $\One_{\Omega}(a)$ refers to the indicator function that takes the value 1 if $a\in\Omega$ and zero otherwise.
Theorem \ref{Th:LK} is fundamental to the classical theories of infinite-divisible laws and L\'evy processes \cite{Sato1994,Steufel2003, Appelbaum2009}.
To further our mathematical understanding of the L\'evy-Khintchine formula (\ref{eq:levykintchine}), we note that
$e^{j a \omega} - 1 - j a \omega \One_{\{|a|<1\}}(a) \sim -\frac{1}{2}a^2 \omega^2$ as $a \rightarrow 0$.
}
This ensures that the integral is convergent even \revision{when the L\'evy measure $V$ is singular} at the origin to the extent allowed by the admissibility condition (\ref{eq:admissibility}). If the L\'evy measure is finite 
%(i.e., $\int_\R V(\dint a)< \infty$) 
or symmetrical (i.e., $V(E)=V(-E)$ for any $E \subset \R$), it is then also possible to use the equivalent, simplified form of L\'evy exponent
\begin{align}
\label{eq:LK2}
f(\omega)=j b_1 \omega - \frac{b_2 \omega^2}{2}+\int_{\RR\backslash\{0\}} \big(e^{j a \omega} - 1\big)\, V(\dint a ) 
\end{align}
with $b_1=b'_1-\int_{0<|a|<1} a V(\dint a)$.
%In this work, we are heavily relying on Fourier analysis and we propose to represent the L\'evy functional by the following equivalent, but non-standard, distributional representation
%\begin{equation}
%\label{eq:levykintchine2}f(u)=j b'_1 u - \frac{b_2 u^2}{2}+\hat v(-u) - \hat v(0)  
%\end{equation}
%where $\hat v=\Fourier\{v\}$ is the generalized Fourier transform of the L\'evy density $v$. Specifically, $v$ is the non-negative generalized function associated with the measure $V$ by way of the identity $\langle \varphi, v \rangle=\lim_{\epsilon \rightarrow 0^+} \int_{|a|>\epsilon} \varphi(a) \dint V(a)$ for any $\varphi \in \mathcal{S}$, which extracts the principal value\footnote{This technical precaution is required because the admissibility condition %$\int_{\R} \min(1,a^2) v(a) \dint a$ 
%allows for a non-integrable singularity at the origin; other than that, the L\'evy density $v(a)\ge0$ is a conventional function of $a$.} of the integral. 
%Our Fourier-domain formula (\ref{eq:levykintchine2}) is much simpler looking than the traditional L\'evy-Khinchine formula.
%The singular aspect of the representation is hidden in the definition of the generalized Fourier transform which naturally takes care of the mathematical technicalities associated with the integration and bias-correction process that is explicit in (\ref{eq:levykintchine}).
\revision{The bottomline is that a particular brand of independent noise process is thereby completely characterized by its L\'evy exponent or, equivalently, its L\'evy triplet $(b_1, b_2, v)$ where $v$ is the so-called {\em L\'evy density} associated with $V$ such that
$$
V(E)=\int_E v(a)\dint a
$$
for any Borel set $E \in\R$.
With this latter convention, the three primary types of white noise
encountered in the signal processing literature are specified as follows:}
\begin{itemize}
\item[1)] Gaussian: $b_1=0, b_2=1, v=0$
$$
f_{\rm Gauss}(\omega)=-\frac{|\omega|^2}{2},$$
\begin{equation}
\label{eq:Ggauss}
\Form_w(\varphi)=e^{-\frac{1}{2} \|\varphi\|_{L_2}^2}.
\end{equation}
\item[2)] Compound Poisson: \quad 
%$ b_1=\lambda \int_{|a|<1} p(a) \dint a, b_2=0, 
$b_1=0, b_2=0$, \revision{$v(a)=\lambda \ p_A(a)$ with $\int_\R
p_A(a) \dint a=\hat p_A(0) =1$,}
$$f_{\rm Poisson}(\omega; \lambda, p_A)=\lambda \int_\R \big(e^{j a \omega} - 1\big) \,p_A(a) \dint a,%=\lambda [\hat p_A(-u) - \hat p_A(0)],
$$
%where $\hat p_a(\omega)=\E\{e^{-j  \omega a}\}$ is the characteristic function of the random amplitude variable $a$.
%The corresponding formulas for the characteristic form of impulsive Poisson noise are
\begin{eqnarray}
\label{eq:Gpoisson}
\Form_w(\varphi)=\exp\left(\lambda \int_\R \int_{\R}    (e^{j a \varphi(t)} - 1) \;p_A(a) \dint a \dint t \right).
%\\&=& \exp\left(\lambda \int_\R \left[\hat p_a\big(-\varphi(t)\big) - \hat p_a(0)\right]\dint t\right) \nonumber
\end{eqnarray}
%which can also be simplified to
%$\displaystyle \Form_w(\varphi)=\exp\left(\lambda \int_\R [\hat p_a(-\varphi(t)) - 1]\dint t\right)$ where $\hat p_a(\omega)=\E\{e^{-j  \omega a}\}$ is the characteristic function of the amplitude variable $a$.
\item[3)] Symmetric alpha-stable (S$\alpha$S): $b_1=0, b_2=0, v(a)=\frac{ C_\alpha }{|a|^{\alpha+1} }$ with $0<\alpha< 2$ and $C_\alpha=\frac{\sin(\frac{\pi \alpha}{2})Ê}{\pi}$ a suitable normalization constant,

$$f_\alpha(\omega)=\frac{-|\omega|^\alpha}{\alpha !},$$
\begin{equation}
\label{eq:alphastable}
\Form_w(\varphi)=e^{-\frac{1}{\alpha!} \|\varphi\|_{L_\alpha}^\alpha}.
\end{equation}
\end{itemize}
The latter follows from the fact that $\frac{-|\omega|^\alpha}{\alpha !}$ is the generalized Fourier transform of $\frac{ C_\alpha }{|t|^{\alpha+1} }$
with the convention that $\alpha !=\Gamma(\alpha+1)$ where $\Gamma$ is Euler's Gamma function \cite{Gelfand-Shilov1964}. 
%This is an interesting example where the first-order bias correction in $u$ that is implicit in formula (\ref{eq:levykintchine}) precisely corrects for the singularity at $a=0$ in the corresponding Fourier integral ($\int_\R v(a) e^{j a u} \dint a$). The effect is analogous to the constant-term substraction in the Poisson functional, which ensures that $f(0)=0$.
%

While none of these noises has a classical interpretation as a random function of $t$, we can at least provide an explicit description of the Poisson noise as a random sequence of Dirac impulses (cf. \cite[Theorem 1]{Unser2011})
$$
w_{\lambda}(t)=\sum_{k} a_k \delta(t-t_k)
$$
where the $t_k$ are random locations that are uniformly distributed over $\R$ with density $\lambda$, and where the weights $a_k$ are i.i.d. random variables with pdf $p_A(a)$.
\subsection{Gaussian versus sparse categorization}
\label{Sec:sparse}
%The generalized white noise processes specified by (\ref{eq:gennoise}) and (\ref{eq:levykintchine2}) are not defined pointwise. 
To get a better understanding of the underlying class of \revision{white
noises $w$}, we propose to probe them through some localized analysis window
$\varphi$, which will yield a conventional i.i.d. random variable
\revision{$X=\langle w, \varphi\rangle$} with some pdf $p_\varphi(x)$.
%
%\subsection{Gaussian vs. sparse categorization}
\label{Sec:sparse}
The most convenient choice is to pick the rectangular analysis window
$\varphi(t)={\rm rect}(t)=\One_{[-\frac{1}{2},\frac{1}{2}]}(t)$ \revision{when 
$\langle w,{\rm rect}\rangle$ is well-defined.}
By using the fact that $e^{j a \omega {\rm rect}(t)}-1=e^{j a  \omega}-1$ for $t\in[-\frac{1}{2},\frac{1}{2}]$, and zero otherwise, we find that the characteristic function of $X$ is
simply given by
$$
\hat p_{\rm rect}(\omega)=\Form_w\left( \omega\cdot {\rm rect}(t)\right)=\exp\left( f(\omega)\right),
$$
which corresponds to the generic (L\'evy-Khinchine) form associated with an infinitely-divisible distribution \cite{Sato1994,Bose2002,Steufel2003}. The above result makes the mapping between generalized white noise processes and classical infinite-divisible (id) laws\footnote{A random variable $X$ with pdf $p_X(x)$ is said to be infinitely divisible (id) if for any $n \in \N^+$ there exist i.i.d. random variables  
$X_{1} , \dots,X_{n}$ with pdf say $p_n(x)$ such that $X=X_{1} + \cdots+X_{n}$ in law.} explicit: The ``canonical" id pdf of $w$, $p_{\rm id}(x)=p_{\rm rect}(x)$, is obtained by observing the noise through a rectangular window. Conversely, given the L\'evy exponent of an id distribution, $f(\omega)=\log \left(\Fourier\{ p_{\rm id}\}(\omega)\right)$, we can specify a corresponding generalized white noise process $w$ via the characteristic form $\Form_w( \varphi)$ by merely substituting the frequency variable $\omega$ by the generic test function $\varphi(t)$, adding an integration over $\R$ and taking the exponential as in (\ref{eq:gennoise}).

%\revision{We will now argue that this class of models allows for a range of behaviors that varies between the purely Gaussian and sparse extremes. In the context of L\'evy processes, these are  often referred to as the diffusive and jump modes. Our understanding of the notion of ``sparsity", which is consistent with the currently dominant paradigm in signal processing, is a concentration of probability around zero and/or a heavy tail behavior of the PDF which implies that most of the ``energy" of a finite sequence of id random variables is contained in a small number of instances. }
\revision{ We note, in passing, that sparsity in signal processing may refer
to two distinct notions.  The first is that of a finite rate of innovation; i.e., a finite (but perhaps random) number of innovations per unit of
time and/or space, which results in a mass at zero in the histogram of
observations.  The second possibility is to have a large, even infinite,
number of innovations, but with the property that a few large innovations
dominate the overall behavior.  In this case the histogram of observations is
distinguished by its `heavy tails'.  (A combination of the two is also
possible, for instance in a compound Poisson process with a heavy-tailed
amplitude distribution.  For such a process one may observe a change of
behavior in passing from one dominant type of sparsity to the other.)  Our
framework permits us to consider both types of sparsity, in the former case
with compound Poisson models and in the latter with heavy-tailed
infinitely-divisible innovations.}

To make our point, we consider two distinct scenarios.

\subsubsection{Finite variance case}
We first assume that the second moment $m_2=\int_{\RR\backslash\{0\}} a^2\, V(da)$ of the L\'evy density $V$ in (\ref{eq:levykintchine}) is finite. 
This allows us to rewrite the classical L\'evy-Khinchine representation as 
$$
f(\omega)=j c_1 \omega - \frac{b_2 \omega^2}{2}+\int_{\R \backslash\{0\}} [e^{j a \omega} - 1 - j a \omega ] \, V(\dint a) 
$$
with $c_1=b_1''+\int_{|a|>1} aV(\dint a)$ and where 
the Poisson part of the functional is now fully compensated.
Indeed, we are guaranteed that the above integral is convergent because $ |e^{j a \omega} - 1 - j \omega a|\lesssim |a \omega |^2$ as $a\rightarrow 0$ and $|e^{j a \omega} - 1 - j \omega a|\sim |a \omega|$ as $a\rightarrow \pm\infty$. 
%$\int_{\R \backslash\{0\}} |a| \dint V(a)<\infty$.
An interesting non-Poisson example of infinitely-divisible probability laws that falls into this category (with non-finite $V$) is 
the Laplace distribution  with L\'evy triplet $(0,0, v(a)=\frac{e^{-|a|}}{|a|}) $ and $p(x)=\frac{1}{2} e^{-|x|}$. This model is particularly relevant for sparse signal processing because it provides a tight connection between L\'evy processes and total variation regularization \cite[Section VI]{Unser2011}.
%There are also many other possibilities (hyperbolic, Gamma, Meixner distributions) which are being used primarily in financial modeling.

Now, if the L\'evy measure is finite $\int_{\R} V(\dint a) =\lambda<\infty$,
the admissibility condition yields $\int_{\RR\backslash\{0\}} a\ V(\dint a)<\infty$,
which allows us to pull the bias correction out of the integral.
The representation then simplifies to \eqref{eq:LK2}.
%\revision{%
%$$
%f(\omega)=j b_1 \omega - \frac{b_2 \omega^2}{2}+\lambda
%\int_{\RR\backslash\{0\}} [e^{j a \omega} - 1]\; 
%    %p_A(a) \dint a,
%    V(\dint a)
%$$
%%with $V(\dint a)=\lambda p_A(a)\dint a$ and $\int_\R p_A(a) \dint a =1$. 
%with $b_1\in\RR$ and $b_2\geq 0$ as defined previously.}
This implies that we can decompose $X$ into the sum of two independent
Gaussian and compound Poisson random variables. The variances of the Gaussian
and Poisson components are $\sigma^2=b_2$ and \revision{$\int_\R a^2 V(\dint
a)$}, respectively. 
The Poisson component is sparse because its
pdf exhibits a mass distribution $e^{-\lambda}\delta(x)$ at the origin, meaning that the chances for a continuous amplitude distribution of getting zero are overwhelmingly higher than any other value, especially for smaller values of $\lambda>0$. It is therefore justifiable to use $0\le e^{-\lambda}<1$ as our Poisson sparsity index. 
%This indicates that the underlying generalized processes are intrinsically sparse, except for the two extreme cases of the representation: 1) pure Gaussian noise with $(b_2>0,\lambda=0)$, and 2) degenerate Poisson noise with $\lambda\rightarrow \infty$ (depending on wether the tail behavior of $p(x)$ is subexponential or not).

\subsubsection{Infinite variance case}
We now turn our attention to the case where the second moment of the L\'evy measure is unbounded, which we like to label as the ``super-sparse" one. To substantiate this claim, we invoke the Ramachandran-Wolfe theorem which states that the $p$th moment $\E\{|X|^p\}$ with $p\inR^+$ of an infinitely divisible distribution  is finite iff. $\int_{|a|>1} |a|^p \ V(\dint a)<\infty$ \cite{Ramachandran1969,Wolfe1971}.
For $p\ge2$, the latter is equivalent to $\int_{\RR\backslash\{0\}} |a|^p\
V(\dint a)<\infty$ because of the \revision{admissibility condition
\eqref{eq:admissibility}}. It follows that the cases that are not covered by
the previous scenario (including the Gaussian + Poisson model) necessarily
give rise to distributions whose moments of order $p$ are unbounded for
$p\ge2$. The prototypical representatives of such heavy tail distributions are
the alpha-stable ones or, by extension, the broad family of infinite divisible
probability laws that are in their domain of attraction. Note that these distributions all fulfill the stringent conditions for $\ell_p$
compressibility\cite{Amini2011}.
\section{Innovation approach to continuous-time stochastic processes}
Specifying a stochastic process through an innovation model (or an equivalent stochastic differential equation) is attractive conceptually, but it presupposes that we can provide an inverse operator (in the form of an integral transform) that transforms the white noise back into the initial stochastic process. This is the reason why we will spend the greater part of our effort investigating suitable inverse operators.

\subsection{Stochastic differential equations}
%Let us consider a linear differential operator $\Lop: \mathcal{S}' \rightarrow \mathcal{S}'$.
Our aim is to define the generalized process with whitening operator  $\Lop$ and L\'evy exponent $f$ as the solution of the stochastic linear differential equation
%\footnote{This equation as well as the following ones are to be interpreted in the distributional sense; that is, $\langle\Lop s, \varphi \rangle=\langle w, \varphi \rangle, \forall \varphi \in \mathcal{S}$; the more common notation $\Lop s(t)=w(t)$ is potentially misleading because $s$ and $w$ are random distributions whose sample values $s(t)$ and $w(t)$ may not always be defined.}
\begin{equation}
\label{eq:operatoreq}
\Lop s=w,
\end{equation}
where $w$ is a white noise process, as described in Section 
\ref{sec:whitenoise}.
This definition is obviously only usable if we can construct an inverse operator $\Top=\Lop^{-1}$ that solves this equation.
For the cases where the inverse is not unique, we will need to select one preferential operator, which is equivalent to imposing specific boundary conditions. We are then able to formally express the stochastic process as a transformed version of a white noise
\begin{eqnarray}
\label{eq:operatorsol}
s=\Lop^{-1}w.
\end{eqnarray}
The requirement for such a solution to be consistent with (\ref{eq:operatoreq}) is that the operator satisfies the right-inverse property $\Lop\Lop^{-1}=\Identity$ over the underlying class of tempered distributions.
By using the adjoint relation $\langle s, \varphi \rangle=\langle \Lop^{-1}w, \varphi \rangle=\langle w, \Lop^{-1\ast}\varphi \rangle$, we can then transfer the action of the operator onto the test function inside the characteristic form and obtain a complete statistical characterization of the so-defined generalized stochastic process
\begin{eqnarray}
\label{eq:formgprocess}
\Form_s(\varphi)=\Form_{\Lop^{-1}w}(\varphi)=\Form_w(\Lop^{-1\ast}\varphi),
\end{eqnarray}
where $\Form_w$ is given by (\ref{eq:gennoise}) (or one of the specific forms
in the list at the end of Section \ref{sec:whitenoise}) and where we are
implicitly requiring that the adjoint $\Lop^{-1\ast}$ is
mathematically well-defined (continuous) over $\mathcal{S}$, and that its
composition with $\Form_w$ is well-defined for all $\phi\in\Spc S$.

In order to realize the above idea mathematically, it is usually easier to proceed backwards: % (Eq. (\ref{eq:formgprocess})): 
one specifies an operator $\Top$ that satisfies the left-inverse property:
$\forall \varphi \in \mathcal{S},\ \Top\Lop^{\ast}\varphi=\varphi$, and that
is continuous (i.e., bounded in a proper topology) over the chosen class of test functions. One then characterizes the adjoint of $\Top$, which, for a given $\phi \in \mathcal{S}$,  is such that
$$\forall \varphi \in \mathcal{S},\quad \langle \Top\varphi, \phi \rangle=\langle \varphi, \Top^\ast \phi \rangle.$$
Finally, one applies a standard limit argument to extend the action of $\Top^\ast=\Lop^{-1}$ over the enlarged class of tempered distribution $\phi \in \mathcal{S}'$ based on the above adjoint relation, which yields the proper distributional definition of the right inverse of $\Lop$ in (\ref{eq:operatorsol}).
\subsection{Inverse operators}
\label{sec:inverse}
Before presenting our general method of solution, we need to identify a suitable set of elementary inverse operators that satisfy the required boundedness conditions.

Our approach relies on the factorization of a differential operator into simple first-order components of the form $(\Dop- \alpha_n \Identity)$ with $\alpha_n \in \C$, which can then be treated separately.
Three possible cases need to be considered.

{\em 1) Causal-stable: ${\rm Re}(\alpha_n)<0$.} This is the classical textbook hypothesis which leads to a causal-stable convolution system. It is well known from linear system theory that the causal Green function of $(\Dop- \alpha_n \Identity)$ is 
the causal exponential function $\rho_{\alpha_n}(t)$ already encountered in the introductory example in Section \ref{Sec:motiv}.
%	$$\rho_{\alpha_n}(t)=u(t) e^{\alpha_n t},$$
%	where $u(t)=\One_{[0,+\infty)}(t)$ is the unit-step (or Heaviside) function.
Clearly, $\rho_{\alpha_n}(t)$ is absolutely integrable (and rapidly-decaying) iff. ${\rm Re}(\alpha_n)<0$. It follows that $(\Dop- \alpha_n \Identity)^{-1} f=\rho_{\alpha_n} \ast f$ with $\rho_{\alpha_n} \in \mathcal{R} \subset L_1$. In particular, this implies that $\Top=(\Dop- \alpha_n \Identity)^{-1}$ specifies a continuous LSI operator on $\mathcal{S}$. The same holds for $\Top^\ast=(\Dop- \alpha_n \Identity)^{-1\ast}$, which is defined as $\Top^\ast f=\rho_{\alpha_n}^\vee \ast f$.

{\em 2) Anti-causal stable: ${\rm Re}(\alpha_n)>0$.} This case is usually excluded because the standard Green function $\rho_{\alpha_n}(t)= \One_+(t) e^{\alpha_n t}$ grows exponentially, meaning that the system does not have a stable causal solution. Yet, it is possible to consider an alternative anti-causal Green function $\rho'_{\alpha_n}(t)=-\rho_{-\alpha_n}^\vee(t)=\rho_{\alpha_n}(t)-e^{\alpha_n t}$, which is unique in the sense that it is the only Green function\footnote{: $\rho$ is a Green functions of $(\Dop- \alpha_n \Identity)$ iff. $(\Dop- \alpha_n \Identity)\rho=\delta$; the complete set of solutions is given  $\rho(t)=\rho_{\alpha_n}(t)+ C e^{\alpha_n t}$ which is the sum of the causal Green function $\rho_{\alpha_n}(t)$ plus an arbitrary exponential component that is in the null space of the operator. } of $(\Dop- \alpha_n \Identity)$ that is Lebesgue-integrable and, by the same token, the proper inverse Fourier transform of $\frac{1}{j \omega-\alpha_n}$ for ${\rm Re}(\alpha_n)>0$.  In this way, we are able to specify an anti-causal inverse filter $(\Dop- \alpha_n \Identity)^{-1} f=\rho'_{\alpha_n} \ast f$ with $\rho'_{\alpha_n} \in \mathcal{R}$ that is $L_p$-stable and $\mathcal{S}$-continuous. In the sequel, we will drop the $'$ superscript with the convention that $\rho_{\alpha}(t)$ systematically refers to the unique Green function of $(\Dop- \alpha \Identity)$ that is rapidly-decay when ${\rm Re}(\alpha)\ne0$.
For now on, we shall therefore use the definition
\begin{eqnarray}
\label{eq:rhoalpha}
\rho_{\alpha}(t)=\left\{
\begin{array}{ll}
 \One_+(t) e^{\alpha t} & \text{if  ${\rm Re}(\alpha)\le 0$} \\
- \One_+(-t) e^{\alpha t} & \text{otherwise.} 
\end{array} \right.
\end{eqnarray}
which also covers the next scenario. 
%which is either causal or anti-causal depending on the polarity of ${\rm Re}(\alpha)$.

{\em 3) Marginally stable: ${\rm Re}(\alpha_n)=0$ or, equivalently, $\alpha_n=j \omega_0$ with $\omega_0 \inR$.} This third case, which is incompatible with the conventional formulation of stationary processes, is most interesting theoretically because it opens the door to important extensions such as L\'evy processes, as we shall see in Section \ref{Sec:propgprocesses}. Here, we will show that marginally-stable systems can be handled within our generalized framework as well, thanks to the introduction of appropriate inverse operators.

%{\em The main difficulty with the third case is the lack of existence of a Green function that is in $L_1$. Concretely, this means that
%$(\Dop- j\omega_0 \Identity)$ does not admit a shift-invariant inverse that is stable in the conventional BIBO sense.
%The difficulty, though, it not insurmountable because the functions that are in the null space of the operator (complex sinusoids and eventually polynomials in the case of $n$-fold iterates) are of slow growth, so that they can be handled by the proposed distributional formalism. We will see that the null space components can be actually be used constructively for adding (non-stationary) trends to the signals in order to meet some prescribed boundary conditions.}

The first natural candidate for $(\Dop- j\omega_0 \Identity)^{-1}$ is the  inverse filter whose
frequency response is
$$
\hat \rho_{j\omega_0}(\omega)= \frac{1 }
       {j(\omega-\omega_0)} + \pi \delta(\omega-\omega_0).
$$
It is a convolution operator whose time-domain definition is 
%impulse response is $\rho_{j\omega_0}(t)=u(t)e^{j \omega_0 t}$. This leads to a convolution-type operator which is defined as follows
\begin{eqnarray}
\label{eq:invshift}
\Iop_{\omega_0}\varphi(t) &=& (\rho_{j\omega_0} \ast \varphi)(t) \nonumber \\
&=& e^{j \omega_0 t}\int_{-\infty}^t e^{-j \omega_0\tau}\varphi(\tau) \dint \tau.
%\\
%&=&\int_{\R}  \displaystyle \hat{\varphi}(\omega) \left(
%   \frac{1 }
%       {j(\omega-\omega_0)} + \pi \delta(\omega-\omega_0)  \right) e^{j \omega t} 
%    \frac{\dint{\omega\;\;}}{2 \pi}
\end{eqnarray}
Its impulse response $\rho_{j\omega_0}(t)$ is causal and compatible with Definition (\ref{eq:rhoalpha}), but not (rapidly) decaying. The adjoint of $\Iop_{\omega_0}$
%where $\displaystyle \hat \varphi(\omega)=\int_{\R}   \varphi(t) e^{-j \omega t} \dint t$ is the Fourier transform of the input function $\varphi \in L_1$. 
%, which is such that $\langle \Iop_{\omega_0}\varphi, \phi \rangle=\langle \varphi,\Iop_{\omega_0}^\ast \phi \rangle$ for all $\varphi, \phi \in \mathcal{S}$, 
is given by
\begin{eqnarray}
\label{eq:invshiftast}
\Iop^\ast_{\omega_0}\varphi(t) &= & (\rho_{j\omega_0}^\vee \ast \varphi)(t)
\nonumber \\
&= &e^{-j \omega_0 t}\int_{t}^{+\infty} e^{j \omega_0\tau}\varphi(\tau) \dint \tau.
\end{eqnarray}
While $\Iop_{\omega_0}\varphi(t)$ and $\Iop^\ast_{\omega_0}\varphi(t)$ are
both well-defined 
%\revision{and continuous} 
when $\varphi \in L_1$, the problem is that these inverse filters are not BIBO stable since their impulse responses, $\rho_{j\omega_0}(t)$ and $\rho^\vee_{j\omega_0}(t)$, are not in $L_1$. In particular, one can easily see that $\Iop_{\omega_0}\varphi$ (resp., $\Iop^\ast_{\omega_0}\varphi$) with $\varphi \in \mathcal{S}$ is generally not in $L_p$ with $1\le p < +\infty$, unless $\hat \varphi(\omega_0)=0$ (resp., $\hat \varphi(-\omega_0)=0$). The conclusion is that $\Iop^\ast_{\omega_0}$ fails to be a bounded operator over the class of test functions $\mathcal{S}$.

This leads us to introduce some ``corrected" version of the adjoint inverse
operator $\Iop^\ast_{\omega_0}$,
\begin{align}
%\begin{eqnarray}
\Iop^\ast_{\omega_0,t_0} \varphi(t) &= \Iop^\ast_{\omega_0}\left\{ \varphi -\hat \varphi(-\omega_0)e^{-j \omega_0 t_0} \delta(\cdot-t_0)\right \}(t) \nonumber \\
&=  \Iop^\ast_{\omega_0} \varphi(t) - \hat \varphi(-\omega_0)e^{-j \omega_0 t_0}\rho_{j\omega_0}^\vee(t-t_0),\ 
\label{eq:invast}
%\end{eqnarray}
\end{align}
where $t_0 \inR$ is a fixed location parameter and where $\hat \varphi(-\omega_0)=\int_\R e^{j \omega_0 t} \varphi(t) \dint t$ is the complex sinusoidal moment associated with the frequency $\omega_0$.
The idea is to correct for the lack of decay of $\Iop^\ast_{\omega_0} \varphi(t)$ as $t \rightarrow -\infty$ by subtracting a properly weighted version of the impulse response of the operator. An equivalent Fourier-based formulation is provided by the formula at the bottom of Table I; the main difference with the corresponding expression for $\Iop_{\omega_0}\varphi$ is the presence of a regularization term in the numerator that prevents the integrant from diverging at $\omega=\omega_0$. The next step is to identify the adjoint of $\Iop^\ast_{\omega_0,t_0}$, which is achieved via the following inner-product manipulation
\begin{align}
\langle \varphi, \Iop^\ast_{\omega_0,t_0}\phi\rangle %&=&\langle \varphi, \Iop^\ast_{\omega_0} \phi - \hat \phi(-\omega_0)e^{j \omega_0 t_0}\rho_{j\omega_0}^\vee(\cdot-t_0)\phi\rangle \\
    &=\langle \varphi, \Iop^\ast_{\omega_0}\phi \rangle - \hat \phi(-\omega_0)e^{-j \omega_0 t_0} \langle \varphi, \rho_{j\omega_0}^\vee(\cdot-t_0)\rangle \onetwocol{& \tag{by linearity}}{\notag}{}\\
%    &=\langle \Iop_{\omega_0} \varphi,\phi \rangle - \hat \phi(-\omega_0)e^{-j \omega_0 t_0} \;\;\Iop_{\omega_0}\varphi(t_0)\right) \tag{using (\ref{eq:invshift})}\\
   &=\langle \Iop_{\omega_0} \varphi,\phi \rangle - \langle e^{j \omega_0 \cdot},\phi \rangle \;e^{-j \omega_0 t_0} \;\;\Iop_{\omega_0}\varphi(t_0) \tag{using (\ref{eq:invshift})}\\
   &=\langle \Iop_{\omega_0} \varphi,\phi \rangle - \langle e^{j \omega_0 (\cdot-t_0)} \Iop_{\omega_0}\varphi(t_0),\phi \rangle. \notag  
   %\\
%    &=\langle \Iop_{\omega_0,t_0} \varphi,\phi \rangle.\tag{by definition}
\end{align}
Since the above is equal to $\langle \Iop_{\omega_0,t_0} \varphi,\phi \rangle$ by definition, we obtain that
\begin{eqnarray}
\label{eq:inv}
\Iop_{\omega_0,t_0} \varphi(t) &=& \Iop_{\omega_0}\varphi (t) - e^{j \omega_0(t-t_0)}\;\Iop_{\omega_0}\varphi(t_0).
\end{eqnarray}
Interestingly, this operator imposes the boundary condition $\Iop_{\omega_0,t_0} \varphi(t_0)=0$ via the substraction of a sinusoidal component that is in the null space of the operator $(\Dop-j \omega_0 \Identity)$, which gives a direct interpretation of the location parameter $t_0$.
Observe that expressions (\ref{eq:invast}) and (\ref{eq:inv}) define linear operators, albeit not shift-invariant ones, in contrast with the classical inverse operators $\Iop_{\omega_0}$ and  $\Iop^\ast_{\omega_0}$.

For analysis purposes, it is convenient to relate the proposed inverse operators to the anti-derivatives corresponding to the case $\omega_0=0$.
To that end, we introduce the modulation operator
$$
\Mop_{\omega_0}\varphi(t)=e^{j\omega_0 t} \varphi(t)
$$
which is a unitary map on $L_2$ with the property that $\Mop^{-1}_{\omega_0}=\Mop_{-\omega_0}$.
\begin{proposition}
The inverse operators defined by (\ref{eq:invshift}), (\ref{eq:invshiftast}), (\ref{eq:inv}), and (\ref{eq:invast}) satisfy the modulation relations
\begin{align*}
\Iop_{\omega_0}\varphi(t)&=\Mop_{\omega_0}\, \Iop_{0}\, \Mop^{-1}_{\omega_0}  \varphi(t), \\
\Iop^\ast_{\omega_0}\varphi(t)&=\Mop^{-1}_{\omega_0} \,\Iop^\ast_{0} \, \Mop_{\omega_0}  \varphi(t),\\
\Iop_{\omega_0,t_0}\varphi(t)&=\Mop_{\omega_0} \,\Iop_{0,t_0}\, \Mop^{-1}_{\omega_0}  \varphi(t), \\
\Iop^\ast_{\omega_0,t_0}\varphi(t)&=\Mop^{-1}_{\omega_0} \,\Iop^\ast_{0,t_0} \, \Mop_{\omega_0}  \varphi(t).
\end{align*}
\end{proposition}
\begin{proof}
These follow from 
the modulation property of the Fourier transform (i.e, $\Fourier\{\Mop_{\omega_0} \varphi  \}(\omega)=\Fourier\{ \varphi  \}(\omega-\omega_0)$) and the observations that $\Iop_{\omega_0}\delta(t)=\rho_{j \omega_0}(t)=\Mop_{\omega_0}\rho_{0}(t)$ and $\Iop^\ast_{\omega_0}\delta(t)=\rho^\vee_{j \omega_0}(t)=\Mop_{-\omega_0}\rho^\vee_{0}(t)$ with
$\rho_0(t)= \One_+(t)$ (the unit step function).
\end{proof}
The important functional property of  $\Iop^\ast_{\omega_0,t_0}$ is that it essentially preserves decay and integrability, while $\Iop_{\omega_0,t_0}$ fully retains signal differentiability. Unfortunately, it is not possible to have the two simultaneously unless $\Iop_{\omega_0}\varphi(t_0)$ and $\hat \varphi(-\omega_0)$ are both zero. 
%The precise mathematical statement about the decay and boundedness of $\Iop^\ast_{\omega_0,t_0}$ is as follows.
\begin{proposition} 
If \revision{$f \in L_{\infty,\alpha}$} with $\alpha>1$, then there exists a constant $C_{t_0}$ such that
\begin{eqnarray*}
|\Iop^\ast_{\omega_0,t_0}f(t)| &\le& C_{t_0} \frac{\|f\|_{\infty,\alpha} }{1 + |t|^{\alpha-1}},
\end{eqnarray*}
which implies that $\Iop^\ast_{\omega_0,t_0} f \in L_{\infty,\alpha-1}$. 
\end{proposition}

{\em Proof}: Since modulation does not affect the decay properties of a function, we can invoke Proposition 1 and concentrate on the investigation of the anti-derivative operator $\Iop^\ast_{0,t_0}$.
Without loss of generality, we can also pick $t_0=0$ and transfer the bound to any other finite value of $t_0$ by adjusting the value of the constant $C_{t_0}$.
Specifically, for $t<0$, we write this inverse operator as
\begin{align*}
\Iop^\ast_{0,0} f(t)&=\Iop^\ast_{0} f(t)-\hat f(0)\\
&=\int_{t}^{+\infty} f(\tau) \dint \tau - \int_{-\infty}^{\infty} f(\tau) \dint \tau \\
&= -\int_{-\infty}^{t} f(\tau) \dint \tau.
\end{align*}
This implies that
\begin{align*} | \Iop^\ast_{0,0} f(t)|= \left|
\int_{-\infty}^{t} f(\tau) \dint \tau \right|& \le  \|f\|_{\infty,
\alpha}\int_{-\infty}^t \frac{1}{1+|\tau|^\alpha} \dint \tau \onetwocol{}{Ê\\ &}\le
\left(\frac{2\alpha}{{\alpha-1}}\right) \frac{\|f\|_{\infty, \alpha} }
{1+|t|^{\alpha-1}}
\end{align*}
\revision{for all $t<0$.}
 For $t>0$, $\Iop^\ast_{0,0} f(t)=\int_{t}^{\infty} f(\tau) \dint
\tau$ so that the above upper bounds remain valid. 
%The final
%result is obtained by combining these two inequalities.
%This implies that\begin{eqnarray*}
%| \Iop^\ast_{0,0} f(t)|=  \left| \int_{-\infty}^{t} f(\tau) \dint \tau \right| \le  \|f\|_{1}
%\end{eqnarray*}
%where $\|f\|_{1}$ is necessarily finite, as a consequence of the embedding relation $L_{\infty,1+\epsilon} \subset L_1$ for any $\epsilon>0$.
%Moreover, for $t<0$, we also have that  
%\begin{eqnarray*}
%|\Iop^\ast_{0,0}f(t)|=\left|\int_{-\infty}^{t} f(\tau) \dint \tau \right|&<& \|f\|_{\infty,\alpha}  \int_{-\infty}^{t} |\tau|^{-\alpha} \dint \tau=\frac{ \|f\|_{\infty,\alpha}}{|t|^{-\alpha+1}}
%\end{eqnarray*}
%For $t>0$, $\Iop^\ast_{0,0} f(t)=\int_{t}^{\infty} f(\tau) \dint \tau$ so that the above upper bounds remain valid. The final result is obtained by combining these two inequalities.
\endproof
The interpretation of the above result is that the inverse operator $\Iop^\ast_{\omega_0,t_0}$ reduces inverse polynomial decay by one order.
%The argument also shows that $\Iop^\ast_{\omega_0,t_0}$ preserves compact support. 
Proposition 2 actually implies that the operator will preserve the rapid decay of the Schwartz functions which are  included in $L_{\infty,\alpha}$ for any $\alpha \in \R^+$. It also guarantees that $\Iop^\ast_{\omega_0,t_0}\varphi$ belongs to $L_{p}$ for any Schwartz function $\varphi$.
However, $\Iop^\ast_{\omega_0,t_0}$ will spoil the global smoothness properties of $\varphi$ because it introduces a discontinuity at $t_0$, unless $\hat \varphi(-\omega_0)$ is zero in which case the output remains in the Schwartz class. This allows us to state the following theorem which summarizes the higher-level part of those results for further reference.
\begin{theorem}
The operator $\Iop^\ast_{\omega_0,t_0}$ defined by (\ref{eq:inv}) is a
continuous linear map \revision{from $\mathcal{R}$ into} $\mathcal{R}$ (the space of bounded functions with rapid decay). Its adjoint $\Iop_{\omega_0,t_0}$ is given by (\ref{eq:invast}) and has the property that $\Iop_{\omega_0,t_0}\varphi(t_0)=0$.
Together, these operators satisfy the complementary left- and right-inverse relations 
\begin{align*}
\left\{
\begin{array}{l}
\Iop^\ast_{\omega_0,t_0}(\Dop-j\omega_0\Identity)^\ast \varphi=\varphi \\
(\Dop-j\omega_0\Identity)\Iop_{\omega_0,t_0} \varphi=\varphi
\end{array}\right.
\end{align*}
for all $\varphi \in \mathcal{S}$.
\end{theorem}

Having a tight control on the action of $\Iop^\ast_{\omega_0,t_0}$ over $\mathcal{S}$ allows us to extend the right-inverse operator $\Iop_{\omega_0,t_0}$ to an appropriate subset of tempered distributions $\phi \in \mathcal{S}'$ according to the rule
$\langle \Iop_{\omega_0,t_0}\phi, \varphi \rangle=\langle \phi, \Iop^\ast_{\omega_0,t_0}\varphi\rangle.$
Our complete set of inverse operators is summarized in Table I together with their equivalent Fourier-based definitions which are also interpretable in the generalized sense of distributions. 
\begin{table*}
\caption{First-order differential operators and their inverses}
\centering
\begin{tabular}{cll}
 \hline \\[-1ex]
$\Lop$ & \hspace{3cm}$\displaystyle \Lop^{-1}f(t)$&Properties of inverse operator\\[2ex]
\hline\\[-1.5ex]
Standard case: $\alpha_n\in \C, {\rm Re}(\alpha_n)\ne0$\\[1ex]
$(\Dop-\alpha_n\Identity)$ & $\displaystyle (\Dop-\alpha_n\Identity)^{-1}f(t) =\int_{\R}   \hat{f}(\omega) \left(
   \frac{1 }
       {j\omega-\alpha_n}  \right)
    e^{j \omega t}\frac{\dint{\omega\;\;}}{2 \pi}$ &$L_p$-stable, LSI, $\mathcal{S}$-continuous \\[4ex]
$(\Dop-\alpha_n\Identity)^\ast$ & $\displaystyle (\Dop^\ast-\alpha_n\Identity)^{-1}f(t) =\int_{\R}   \hat{f}(\omega) \left(
   \frac{1 }
       {-j\omega-\alpha_n}  \right)
    e^{j \omega t}\frac{\dint{\omega\;\;}}{2 \pi}$ &$L_p$-stable, LSI, $\mathcal{S}$-continuous\\
\\[-1.5ex]
Critical case: $\alpha_n=j\omega_0, \omega_0 \inR$\\[1ex]
$(\Dop-j\omega_0\Identity)$ & $\displaystyle \Iop_{\omega_0}f(t) =\int_{\R}   \hat{f}(\omega) \left(
   \frac{1 }
       {j(\omega-\omega_0)} + \pi \delta(\omega-\omega_0) \right)
    e^{j \omega t}\frac{\dint{\omega\;\;}}{2 \pi}$ &Causal, LSI\\[5ex]
 & $\displaystyle \Iop_{\omega_0,t_0}f(t) =\int_{\R}   \hat{f}(\omega) \left(
   \frac{e^{j \omega t}-e^{j \omega_0(t-t_0)} e^{j \omega t_0}}
       {j(\omega-\omega_0)} \right)
    \frac{\dint{\omega\;\;}}{2 \pi}
$ & Output vanishes at $t=t_0$ \\[5ex]
$(\Dop-j\omega_0\Identity)^\ast$ & $\displaystyle \Iop^\ast_{\omega_0}f(t) =\int_{\R}   \left(
   \frac{ \hat{f}(\omega) }
       {-j(\omega+\omega_0)} +  \hat{f}(-\omega_0)\pi \delta(\omega+\omega_0) \right)
    e^{j \omega t}\frac{\dint{\omega\;\;}}{2 \pi}$ & Anti-causal, LSI\\[5ex]
 & $\displaystyle \Iop^\ast_{\omega_0,t_0}f(t) =\int_{\R}    \left(
   \frac{\hat{f}(\omega)-\hat f(-\omega_0) e^{-j (\omega+\omega_0)t_0} }
       {-j(\omega+\omega_0)} \right)e^{j \omega t} 
    \frac{\dint{\omega\;\;}}{2 \pi}$ & $L_p$-stable and decay preserving
 \\[5ex]
 \hline
\end{tabular}
\end{table*}

\subsection{Solution of generic stochastic differential equation}
\label{sec:solutionODE}
We now have all the elements to solve the generic stochastic linear differential equation
\begin{eqnarray}
 \sum_{n=1}^N a_n \Dop^n s=  \sum_{m=1}^M b_m \Dop^m w
\label{eq:ODE}
\end{eqnarray}
where the $a_n$ and $b_m$ are arbitrary complex coefficients with the
normalization constraint $a_N=1$. While this reminds us of the textbook
formula of an ordinary $N$th-order differential system, the non-standard
aspect \revision{in \eqref{eq:ODE}} is that the driving term is a white noise process $w$, which is generally not defined pointwise, and that we are not imposing any stability constraint. Eq. (\ref{eq:ODE}) thus covers the general case 
(\ref{eq:operatoreq}) where $\Lop$ is a shift-invariant operator
with the rational transfer function
\begin{align}
\hat L(\omega)&=\frac{(j\omega)^N+a_{N-1}(j\omega)^{N-1}+\cdots+a_1(j\omega) +a_0} {b_{M}(j\omega)^{M}+\cdots+b_1(j\omega) +b_0}\onetwocol{}{\nonumber \\ &}=\frac{P_N(j\omega)}{Q_M(j\omega)}.
\label{eq:frequencyrep}
\end{align}
%where the interpretation  in the general sense of distributions.
The poles of the system, which are the roots of the characteristic polynomial $P_N(\s)=\s^N+a_{N-1}\s^{n-1}+\cdots+ a_0$ with Laplace variable $\s\in\C$, are denoted by $\{\alpha_n\}_{n=1}^N$. While we are not imposing any restriction on their locus in the complex plane, we are adopting a special ordering where the purely imaginary roots (if present) are coming last. This allows us to factorize the numerator of (\ref{eq:frequencyrep}) as 
\begin{align}
P_N(j\omega)%&=&(j\omega)^N+a_{N-1}(j\omega)^{N-1}+\cdots+a_1(j\omega) +a_0 \nonumber \\
&=\prod_{n=1}^N(j\omega - \alpha_n)  \onetwocol{}{\nonumber\\&}=\left(\prod_{n=1}^{N-n_0}(j\omega - \alpha_n)\right)\,\left( \prod_{m=1}^{n_0}(j\omega - j \omega_m)\right)
\end{align}
with  $\alpha_{N-n_0+m}=j \omega_m$, \revision{$1\leq m\leq n_0$,} where $n_0$ is the 
number of purely-imaginary poles.
The operator counterpart of this last equation is the decomposition
\begin{align*}
P_N(\Dop)=\underbrace{(\Dop - \alpha_1\Identity) \cdots (\Dop - \alpha_{N-n_0}\Identity)}_{\mbox{regular part}} \onetwocol{\,}{ \hspace{0cm}\\ \circ}\underbrace{(\Dop - j\omega_1 \Identity) \cdots (\Dop - j\omega_{n_0} \Identity)}_{\mbox{critical part}}
\end{align*}
which involves a cascade of elementary first-order components. By applying the proper sequence of right-inverse operators from Table I, we can then formally solve the system as in (\ref{eq:operatorsol}). The resulting inverse operator is
\begin{align}
\Lop^{-1}&=\underbrace{\Iop_{\omega_{n_0}, t_{n_0}} \cdots \Iop_{\omega_{1}, t_1} }_{\mbox{shift-variant}}\Top_{\rm LSI}
\label{eq:invODE}
\end{align}
with $$\Top_{\rm LSI}=(\Dop-\alpha_{N-n_0}\Identity)^{-1} \cdots (\Dop-\alpha_1\Identity)^{-1} Q_M(\Dop),$$
which imposes the $n_0$ boundary conditions
\begin{eqnarray}
\label{eq:boundary}
\left\{
\begin{array}{rcl}
\left. s(t)\right|_{t=t_{n_0}}&=&0  \\
\left. (\Dop-j \omega_{n_0}\Identity)s(t)\right|_{t=t_{n_0-1}}&=&0\\
&\vdots&\\
\left.  (\Dop-j \omega_{2}\Identity)\cdots(\Dop-j \omega_{n_0}\Identity) s(t)\right|_{t=t_1}&=&0.
\end{array}
\right.
\end{eqnarray}
The corresponding adjoint operator is given by
\begin{eqnarray}
\Lop^{-1\ast}=\Top_{\rm LSI}^\ast \underbrace{\Iop^\ast_{\omega_{1}, t_1} \cdots \Iop^\ast_{\omega_{n_0}, t_{n_0}}}_{\mbox{shift-variant}},
\label{eq:invODEast}
\end{eqnarray}
and is guaranteed to be a continuous linear mapping from $\mathcal{S}$ into $\mathcal{R}$ by Theorem 1, the key point being that each of the component operators preserves the rapid decay of the test function to which it is applied. The last step is to substitute the explicit form (\ref{eq:invODEast})
of $\Lop^{-1\ast}$ into (\ref{eq:formgprocess}) \revision{with a $\Form_w$ 
that is well-defined on $\Spc R$}, which yields the characteristic form of the stochastic process $s$ defined by (\ref{eq:ODE}) subject to the boundary conditions  (\ref{eq:boundary}).

We close this section with a comment about commutativity: while the order of application of the operators $Q_M(\Dop)$ and $(\Dop-\alpha_n\Identity)^{-1}$ in the LSI part of (\ref{eq:invODE}) is immaterial (thanks to the commutativity of convolution), it is not so for the inverse operators $\Iop_{\omega_m,t_0}$ that appear in the ``shift-variant" part of the decomposition. The latter do not commute and their order of application is tightly linked to the boundary conditions.
\section{Sparse stochastic processes}
\label{Sec:propgprocesses}
This section is devoted to the characterization and investigation of the properties of the broad family of stochastic processes specified by the innovation model (\ref{eq:operatoreq}) where $\Lop$ is LSI.
It covers the non-Gaussian stationary processes (\ref{Sec:stationary}), which are generated by conventional analog filtering of a sparse innovation, as well as the whole class of processes that are solution of the (possibly unstable) differential equation \eqref{eq:ODE} with a L\'evy noise excitation (\ref{Sec:Glevy}). The latter category constitutes the higher-order generalization of the classical L\'evy processes, which are non-stationary. 
%One argument for calling these processes sparse is that they admit an expansion in a wavelet-like basis that is (nearly) decoupled and intrinsically sparse, as shown in Section \ref{Sec:wav}.

We have just addressed the fundamental issue of the solvability of the operator equation $\Lop s=w$.
% has just been addressed in Section \ref{sec:solutionODE}. 
The only missing ingredient is that one needs to ensure that the formal solution $s=\Lop^{-1}w$ is a bona fide generalized stochastic process. 
The answer, of course, is dependent upon whether or not we are able to exhibit an (adjoint) inverse operator $\Top=\Lop^{-1\ast}$ that is sufficiently well-behaved for the resulting characteristic form $\Form_s(\varphi)=\Form_w(\Top\varphi)$ to satisfy the sufficient conditions (continuity, positive-definiteness, and normalization) for existence, as stated in the Minlos-Bochner theorem (Theorem \ref{Th:Minlos}). To that end, we shall rely on the following result whose proof is given in Appendix II.

\begin{theorem}[Admissibility]
\label{th:admissibility}
Let $f$ is a valid L\'evy exponent and $\Top$ is an operator acting on $\varphi \in \mathcal{S}$ 
such that
any one of the conditions below is met:
\begin{enumerate}
\item $\Top$ is a continuous linear map from $\mathcal{S}$ into
itself, \item $\Top$ is a continuous linear map from $\mathcal{S}$
into $L_p$ and the L\'evy exponent $f$ is {\em $p$-admissible} in the sense that $|f(u)|+|u|\cdot |f'(u)|\le
C|u|^p$ for all $u\inR$, where $1\le p<\infty$ and $C$ is
a positive constant.
\end{enumerate}
Then, $\Form_s(\varphi)=$ $\exp\left(\int_{\R} f\big(\Top\varphi(t)\big) \dint
t\right)$ is a continuous, positive-definite functional on $\mathcal{S}$ such that $\Form_s(0)=1$.%, and hence to define a generalized stochastic process over $\mathcal{S}'$,
\end{theorem}
%Let $\Form_s(\varphi)=\exp\left( \int_{\R} f\big(\Top\varphi\big) \dint t\right)$ where the scalar function $f(u)$ is of the generic L\'evy-Khintchine form (\ref{eq:levykintchine}) and where $\Top$ is an operator acting on $\varphi \in \mathcal{S}$. Then, $\Form_s(\varphi)$ is a continuous, positive-definite functional on $\mathcal{S}$ such that $\Form_s(0)=1$,
%%, and hence to define a generalized stochastic process over $\mathcal{S}'$, 
%provided that any one of the conditions below is met
%\begin{enumerate}
%\item $\Top$ is a continuous linear map from $\mathcal{S}$ into itself, 
%\item $\Top$ is a continuous linear map from $\mathcal{S}$ into $L_p$ and $f(u)$ is such that $|f(u)|+|u|\cdot |f'(u)|\le C|u|^p$ for some $p\ge1$.
%\end{enumerate}
%\end{theorem}

\subsection{Non-Gaussian stationary processes}
\label{Sec:stationary}
The simplest scenario is when $\Lop^{-1}$ is LSI and can be decomposed into a cascade of BIBO-stable and ordinary differential operators. If the BIBO-stable part is rapidly-decreasing, then $\Lop^{-1}$ is guaranteed to be $\mathcal{S}$-continuous. 
In particular, this covers the case of an $N$th-order differential system without any pole on the imaginary axis, as justified by our analysis in Section \ref{sec:solutionODE}.

\begin{proposition} [Generalized stationary processes]
\label{Prop:gstationary}
Let $\Lop^{-1}$ (the right-inverse of some operator $\Lop$) be a
$\mathcal{S}$-continuous convolution operator characterized by its impulse
response $\rho_\Lop=\Lop^{-1}\delta$. Then, the generalized stochastic
processes that are defined by \revision{$\Form_s(\varphi)=\exp\left( \int_{\R}
f\big(\rho_\Lop^\vee \ast \varphi(t)\big) \dint t\right)$} where $f(u)$ is of the generic form (\ref{eq:levykintchine}) are {\em stationary} and well-defined solutions of the operator equation (\ref{eq:operatoreq}) driven by some corresponding innovation process $w$.
\end{proposition} 
\begin{proof}
The fact that these generalized processes are well-defined is a direct
consequence of the Minlos-Bochner Theorem since $\Lop^{-1\ast}$ (the
convolution with $\rho_\Lop^\vee$) satisfies the first admissibility
condition in Theorem \ref{th:admissibility}. The stationarity property is equivalent to $\Form_s(\varphi)=\Form_s(\varphi(\cdot-t_0))$ for all $t_0 \inR$; it is established by simple change of variable in the inner integral using the basic shift-invariance property of convolution; i.e., $\left(\rho_\Lop^\vee \ast \varphi(\cdot-t_0)\right)(t)=(\rho_\Lop^\vee \ast \varphi)(t-t_0)$.
\end{proof}

The above characterization is not only remarkably concise, but also quite
general. It extends the traditional theory of stationary Gaussian processes,
which corresponds to the choice $f(u)=-\frac{\sigma_0^2}{2} u^2$.  The
Gaussian case results in the simplified form \revision{$\int_\R f(\Lop^{-1\ast}
\varphi(t)) \dint t =-\frac{\sigma_0^2}{2}\|\rho_\Lop^\vee \ast
\varphi\|^2_{L_2}=-\frac{1}{4 \pi}\int_\R \Phi_s(\omega)|\hat
\varphi(\omega)|^2\dint \omega$} (using Parseval's identity) where $\Phi_s(\omega)=\frac{\sigma_0^2}{|\hat L(-\omega)|^2}$ is the spectral power density that is associated with the innovation model. The interest here is that we get access to a much broader family of non-Gaussian processes (e.g., generalized Poisson or alpha-stable) with matched spectral properties since they share the same whitening operator $\Lop$.

The characteristic form condenses all the statistical information about the process. For instance, by setting $\varphi=\omega\delta(\cdot-t_0)$, we can explicitly determine
$\Form_s(\varphi)=\E\{ e^{j\langle s, \varphi\rangle} \}=\E\{ e^{j\omega
s(t_0)} \}=\Fourier\{p\big(s(t_0)\big)\}(-\omega)$, which yields the
characteristic function of the first-order probability density,
$p(s(t_0)) = p(s)$, of the sample values of the process. In the present stationary scenario, we find that $p(s)=\Fourier^{-1}\{\exp\left( \int_{\R} f\big(-\omega \rho_\Lop(t)\big) \dint t\right)\}(s)$, which requires the evaluation of an integral followed by an inverse Fourier transform. While this type of calculation is only tractable analytically in special cases, it may be performed numerically with the help of the FFT. Higher-order density functions are accessible as well as at the cost of some multi-dimensional inverse Fourier transforms. The same applies for moments which can be obtained through a simpler differentiation process, as exemplified in Section \ref{sec:corr}.

\subsection{%Solutions of $N$th-order SDE and 
Generalized L\'evy processes}
\label{Sec:Glevy}
The further reaching aspect of the present formulation is that it is also applicable to the characterization of non-stationary processes such as Brownian motion and L\'evy processes, which are usually treated separately from the stationary ones, and that it naturally leads to the identification of a whole variety of higher-order extensions.
The commonality is that these non-stationary processes can all be derived as solutions of an (unstable) $N$th-order differential equation with some poles on the imaginary axis. This corresponds to the setting in Section \ref{sec:solutionODE}
 with $n_0>0$.

\begin{proposition} [Generalized $N$th-order L\'evy processes]
\label{prop:nonstat}
Let $\Lop^{-1}$  (the right-inverse of an $N$th-order differential operator
$\Lop$) be specified by (\ref{eq:invODE}) with at least one
non-shift-invariant factor $\Iop_{\omega_1,t_1}$. Then, the generalized
stochastic processes that are defined by
\revision{$\Form_s(\varphi)=\exp\left( \int_{\R} f\big(\Lop^{-1*}
\varphi(t)\big) \dint t\right)$}, where $f(u)$ is of the generic form (\ref{eq:levykintchine}) subject to the constraint $|f(u)|+|u|\cdot |f'(u)|\le C|u|^p$ for some $p\ge1$, are well-defined solutions of the stochastic differential equation (\ref{eq:ODE}) driven by some corresponding 
L\'evy white noise $w$. These processes satisfy the boundary conditions (\ref{eq:boundary}) and are {\em non-stationary}.
\end{proposition}

Note that the $p$-admissibility condition on the L\'evy exponent $f$ is satisfied by the great majority of the members of the L\'evy-Kintchine family. For
instance in the \revision{compound} Poisson case, we can show that $|u| \cdot
|f'(u)|\le \lambda |u|\; \E\{ |A|\}$ and $f(u)\le \lambda |u|
\;\E\{|A|\}$ by using the fact $|e^{jx}-1|\le |x|$; this implies that the
bound in Theorem 3 with $p=1$ is always satisfied provided that the first
(absolute) moment of the amplitude pdf $p_A(a)$ in
\eqref{eq:Gpoisson} is finite. The only cases we are aware of that do not
fulfill the condition are the alpha-stable noises with $0<\alpha<1$, which are
notorious for their exotic behavior.

\begin{proof}
The result is a direct consequence of the analysis in Section \ref{sec:solutionODE}---in particular, Eqs. (\ref{eq:invODE})-(\ref{eq:invODEast})---and Proposition 2.
The latter implies that $\Lop^{-1*}\varphi$ is bounded in all $L_{\infty,m}$ norms with $m\ge1$.
% where $n_0$ is the number of purely imaginary poles, as specified earlier. 
Since $\mathcal{S} \subset L_{\infty,m} \subset L_p$ and the Schwartz topology is the strongest in this chain,  we can infer that $\Lop^{-1*}$ is a continuous operator from $\mathcal{S}$ onto any of the $L_p$ spaces with $p\ge1$. The existence claim then follows from the combination of Theorem \ref{th:admissibility} and Minlos-Bochner. Since $\Lop^{-1*}\varphi$ is not shift-invariant, there is no chance for these processes to be stationary, not to mention the fact that they fulfill the boundary conditions (\ref{eq:boundary}).
\end{proof}

Conceptually, we like to view the generalized stochastic processes of Proposition \ref{prop:nonstat} as ``adjusted" versions of the stationary ones that include some additional sinusoidal (or polynomial) trends. While the generation mechanism of these trends is random, there is a deterministic aspect to it because it imposes the boundary conditions (\ref{eq:boundary}) at $t_1, \cdots, t_{n_0}$. The class of such processes is actually quite rich and the formalism surprisingly powerful. We shall illustrate the use of Proposition \ref{prop:nonstat} in Section V with the simplest possible operator $\Lop=\Dop$ which will gets us back to Brownian motion and the celebrated family of L\'evy processes. We shall also show how the well-known properties of L\'evy processes can be readily deduced from their characteristic form.

\subsection{Moments and correlation}
 \label{sec:corr}
The covariance form of a generalized (complex-valued) process $s$ is defined as:
$$
\Corr_s(\varphi_1, \varphi_2)= \E\{\langle s, \varphi_1\rangle \cdot  \overline{\langle s, \varphi_2\rangle}\}.
$$
where $\overline{\langle s, \varphi_2\rangle}=\langle s, \varphi_2\rangle$ when $s$ is real-valued. Thanks to the moment generating properties of the Fourier transform, this functional can be calculated from the characteristic form $\Form_s(\varphi)$ as
\begin{equation}
\label{eq:covarfromcharacteristicform}
\Corr_s(\varphi_1,\varphi_2)=(-j)^2 \left.\frac{\partial^2 \Form_s(\omega_1\varphi_1+\omega_2\varphi_2)}{\partial \omega_1 \partial \omega_2}\right|_{\omega_1=0,\omega_2=0},
\end{equation}
where we are implicitly assuming that the required partial derivative of the characteristic functional exists. The autocorrelation of the process is then obtained by making the formal substitution $\varphi_1=\delta(\cdot-t_1)$ and $\varphi_2=\delta(\cdot-t_2)$:
$$
R_s(t_1,t_2)=\E\{s(t_1) s(t_2)\}=\Corr_s\left(\delta(\cdot-t_1), \delta(\cdot-t_2)\right).
$$
Alternatively, we can also retrieve the autocorrelation function by invoking the kernel theorem: $\Corr_s(\varphi_1,\varphi_2)=\int_{\R^2} R_s(t_1,t_2) \varphi_1(t_1) \varphi(t_2)\dint t_1 \dint t_2$.

%Of course, the latter only makes sense if the process has a proper pointwise interpretation which, at the least, requires continuity. 
The concept also generalizes for the calculation of the higher-order correlation form\footnote{For simplicity, we are only giving the formula for a real-valued process.}
\begin{align*}\begin{split}\E\{\langle s, \varphi_1\rangle\cdot\langle s, \varphi_2\rangle \cdots  \langle s, \varphi_N\rangle\}=\onetwocol{}{\hspace*{3cm}\\}
(-j)^N \left.\frac{\partial^N \Form_s(\omega_1\varphi_1+\cdots+\omega_N\varphi_N)}{\partial \omega_1 \cdots \partial \omega_N}\right|_{\omega_1=0,\cdots,\omega_N=0}
\end{split}
\end{align*}
which provides the basis for the determination of higher-order moments and cumulants.

Here, we concentrate on the calculation of the second-order moments, which happen to be independent upon the specific type of noise.
For the cases where the covariance is defined and finite,
%exceptional case of the alpha-stable processes with $\alpha<2$ whose second-order moments are not defined,
it is not hard to show
%\footnote{The formulation excludes the  exceptional case of the alpha-stable processes with $\alpha<2$ since the 2nd-derivative of $|u|^\alpha$  in (-) is not defined at $u=0$, which we take as a confirmation of the well-known fact that the  variance of such distributions is unbounded}
that the generic covariance form  of the white noise processes defined in Section \ref{sec:whitenoise} is $$
\Corr_w(\varphi_1, \varphi_2)= \sigma^2_0 \;\langle \varphi_1, \varphi_2\rangle,
$$
where $ \sigma^2_0$ is a suitable normalization constant that depends on the
noise parameters $(b_1, b_2, v)$ in
\eqref{eq:admissibility}--\eqref{eq:Gpoisson}.  We then perform the usual adjoint manipulation to transfer the above formula to the filtered version $s=\Lop^{-1}w$ of such a noise process. 
\begin{property}[Generalized correlation]
\label{Prop:correlation}
The covariance form of the generalized stochastic process whose characteristic form is $\Form_s(\varphi)=\Form_w(\Lop^{-1\ast}\varphi)$ where $\Form_w$ is a white noise functional is given by
$$
\Corr_s(\varphi_1, \varphi_2)= \sigma^2_0 \;\langle \Lop^{-1\ast}\varphi_1, \overline{\Lop^{-1\ast}\varphi_2}\rangle=\sigma^2_0 \; \langle \overline{\Lop^{-1}}\Lop^{-1\ast}\varphi_1, \varphi_2\rangle,
$$
and corresponds to the correlation function 
$$
R_s(t_1,t_2)=\E\{s(t_1)  \cdot  \overline{s(t_2)}\}= \sigma^2_0 \; \langle \overline{\Lop^{-1}}\Lop^{-1\ast}\delta(\cdot-t_1), \delta(\cdot-t_2)\rangle.
$$
\end{property}

The latter characterization requires the determination of the impulse response of $\overline{\Lop^{-1}}\Lop^{-1\ast}$. 
In particular, when $\Lop^{-1}$ is LSI with convolution kernel $\rho_\Lop \in L_1$, we get that
\begin{align*}R_s(t_1,t_2)&= \sigma^2_0\; \overline{\Lop^{-1}} \Lop^{-1\ast}\delta(t_2-t_1)= r_s(t_2-t_1)\onetwocol{}{\\ &}=\sigma^2_0 \; (\overline{\rho}_\Lop \ast \rho_\Lop^\vee) (t_2 -t_1),
\end{align*}
which confirms that the underlying process is wide-sense stationary. Since the autocorrelation function $r_s(\tau)$ is integrable, we also have a one-to-one correspondence with the traditional notion of power spectrum: $\Phi_s(\omega)=\Fourier\{r_s \}(\omega)=\frac{\sigma_0^2}{|\hat L(-\omega)|^2}$,
where $\hat L(\omega)$ is the frequency response of the whitening operator $\Lop$.

The determination of the correlation function for the non-stationary processes associated with the unstable versions of (\ref{eq:ODE}) is more involved. 
We shall see in \cite{Unser2012b} that it can be bypassed if, instead of $s(t)$, we consider the generalized increment process $s_{\rm d}(t)=\Lop_{\rm d} s(t)$ 
where $\Lop_{\rm d}$ is a discrete version (finite-difference type operator) of the whitening operator $\Lop$.

\subsection{Sparsification in a wavelet-like basis}
\label{Sec:wav}
The implicit assumption for the next properties is that we have a wavelet-like basis $\{\psi_{i,k}\}_{i\inZ, k\inZ}$ available that is matched to the operator $\Lop$. Specifically, the basis functions $\psi_{i,k}(t)=\psi_i(t-2^i k)$ with scale and location indices $(i,k)$ are translated versions of some normalized reference wavelet $\psi_i=\Lop^\ast \phi_i$ where $\phi_i$ is an appropriate scale-dependent smoothing kernel. It turns out that such operator-like wavelets can be constructed for the whole class of ordinary differential operators considered in this paper \cite{Khalidov2006}. They can be specified to be orthogonal and/or compactly supported (cf. examples in Fig. 2). In the case of the classical Haar wavelet, we have that $\psi_{\rm Haar}=\Dop \phi_i$ where the smoothing kernels $\phi_i \propto \phi_0(t/2^{i})$ are rescaled versions of a triangle function (B-spline of degree $1$). The latter dilation property follows from the fact that the derivative operator $\Dop$ commutes with scaling.

We note that the determination of the wavelet coefficients $v_i[k]=\langle s, \psi_{i,k}\rangle$ of the random signal $s$
at a given scale $i$ is equivalent to correlating the signal with the wavelet $\psi_i$ (continuous wavelet transform) and sampling thereafter. The goods news is that this has a stationarizing and decoupling effect.

\begin{property}[Wavelet-domain probability laws]
\label{Prop:wav1}
Let $v_i(t)=\langle s, \psi_i(\cdot -t)\rangle$ with $\psi_i=\Lop^\ast\phi_i$ be the $i$th channel of the continuous wavelet transform  of
a generalized (stationary or non-stationary)  L\'evy process $s$ with whitening operator $\Lop$ and $p$-admissible L\'evy exponent $f$.
Then,  $v_i(t)$ is a generalized stationary process with characteristic functional
$\Form_{v_i}(\varphi)=\Form_w(\phi_i \ast \varphi)$ where $\Form_w$ is defined by \eqref{eq:gennoise}. Moreover, the characteristic function of the (discrete) wavelet coefficient $v_i[k]=v_i(2^i k)$---that is, the Fourier transform of the pdf $p_{v_i}(v)$---%which is independent upon the location $k$ (or $t$)
is given by $\hat p_{v_i}(\omega)=\Form_w(\omega \phi_i)=e^{f_i(\omega)}$ and is infinitely divisible with modified L\'evy exponent
$$
f_i(\omega)=\int_\R f\big(\omega\phi_i(t)\big) \dint t.
$$
\end{property} 
\begin{proof}
Recalling that $s=\Lop^{-1}w$%and $\psi_i=\Lop^\ast\phi_i$
, we get
\begin{align*}
v_i(t)&=\langle s, \psi_i(\cdot -t)\rangle=\langle \Lop^{-1}w, \Lop^\ast\phi_i(\cdot-t)\rangle
%\tag{by duality} 
\\
&=\langle w, \Lop^{-1\ast}\Lop^\ast \phi_i(\cdot-t)\rangle=\big(\phi_i^\vee \ast w\big)(t) \nonumber
\end{align*} 
where we have used the fact that
$\Lop^{-1\ast}$ is a valid (continuous) left-inverse of $\Lop^\ast$. The wavelet smoothing kernel 
$\phi_i\in\mathcal{R}$ has rapid decay (e.g., compactly-support or, at worst, exponential decay); this allows us to invoke Proposition \ref{Prop:gstationary}
to prove the first part.
%, owing to the fact that $v_i[k]=\left.v_i(t)\right|_{t=2^i k}$.

As for the second part, we start from the definition of the characteristic function:\begin{align}
\hat p_{v_i}(\omega)&=\Exp\{e^{j\omega v_i}\}=\Exp\{e^{j\omega \langle s, \psi_{i,k}\rangle}\}=\Exp\{e^{ j\langle s, \omega\psi_{i}\rangle}\} \tag{ by stationarity}\\
&=\Form_s(\omega\psi_{i})=\Form_w(\Lop^{-1\ast}\Lop^\ast \phi_i\omega) %\tag{by definition and proper substitution}
\nonumber \\
&=\Form_w(\omega \phi_i)=\exp\left(\int_\R f\big(\omega \phi_i(t)\big)\dint t \right) \nonumber
\end{align}
where we have used the left-inverse property of $\Lop^{-1\ast}$ and the expression of the L\'evy noise functional. The result then follows by identification.
\footnote
{A technical remark is in order here: the substitution of a non-smooth function such as $\phi_i \in \mathcal{R}$ in the characteristic noise functional $\Form_w$ is legitimate provided that the domain of continuity of the functional can be extended from $\mathcal{S}$ to $\mathcal{R}$. This is no problem when $f$ is $p$-admissible since we can readily adapt the proof of Theorem \ref{th:admissibility} to show that $\Form_w$ 
is a continuous, positive-define functional over $L_p(\R)$, which is a much larger space (and with a weaker topology) than both  $\mathcal{S}$ and $\mathcal{R}$.
%Indeed, we have the inclusion $\mathcal{S}\subset\mathcal{R}\subset L_p(\R)$ for any $p>0$. 
}
\end{proof}
We determine the joint characteristic function of any two wavelet coefficients $Y_1=\langle s, \psi_{i_1,k_1}\rangle$ and $Y_2=\langle s, \psi_{i_2,k_2}\rangle$
with indices $(i_1,k_1)$ and $(i_2,k_2)$ using a similar technique.
\begin{property}[Wavelet dependencies]
\label{Prop:joint}
The joint characteristic function of the wavelet coefficients $Y_1=v_{i_1}[k_1]=\langle s, \psi_{i_1,k_1}\rangle$ and $Y_2=v_{i_2}[k_2]=\langle s, \psi_{i_2,k_2}\rangle$ of the generalized stochastic process $s$ in Property \ref{Prop:wav1} is given by$$
\hat p_{Y_1,Y_2}(\omega_1,\omega_2)=\exp\left(\int_\R f\big(\omega_1 \phi_{i_1}(t-2^{i_1}k_1)+\omega_2 \phi_{i_2}(t-2^{i_2}{k_2})\big)\dint t \right)
$$
where $f$ is the L\'evy exponent of the innovation process $w$. The coefficients are independent if the kernels $\phi_{i_1}(t-2^{i_1}k_1)$ and 
$\phi_{i_2}(t-2^{i_2}{k_2})$ have disjoint support; their correlation is given by
$$
\Exp\{Y_1Y_2\}=\sigma_0^2\; \langle  \phi_{i_1}(\cdot-2^{i_1}k_1), \phi_{i_2}(\cdot-2^{i_2}{k_2})\rangle.
$$
under the assumption that the variance $\sigma_0^2$ of $w$ is finite.
\end{property} 
\begin{proof}
The first formula is obtained by substitution of $\varphi=\omega_1\psi_{i_1,k_1}+\omega_2\psi_{i_2,k_2}$ in %the characteristic functional of the process
$\Exp\{e^{j \langle s, \varphi\rangle} \}=\Form_w(\Lop^{-1\ast}\varphi)$, and simplification using the left-inverse property of $\Lop^{-1\ast}$.
The statement about independence follows from the exponential nature of the characteristic function and the property that $f(0)=0$, which allows for the factorization of the characteristic function when the support of the kernels are distinct (independence of the noise at every point). The correlation formula is obtained by direct application of the first result in Property \ref{Prop:correlation}
with $\varphi_1= \psi_{i_1,k_1}=\Lop^\ast \phi_{i_1}(\cdot-2^{i_1}k_1)$ and $\varphi_2= \psi_{i_2,k_2}=\Lop^\ast \phi_{i_2}(\cdot-2^{i_2}k_2)$.
\end{proof}

These results provide a complete characterization of the statistical distribution of sparse stochastic processes in some
matched wavelet domain. They also indicate that the representation is intrinsically sparse since the
transformed-domain statistics are infinitely divisible. Practically, this translates into the wavelet domain pdfs being heavier tailed than a Gaussian (unless the process is Gaussian) (cf. argumentation in Section \ref{Sec:sparse}).

To make matters more explicit, we consider the case where the innovation process is S$\alpha$S. The application of Property \ref{Prop:wav1} with $f(\omega)=-\frac{|\omega|^\alpha}{\alpha !}$ yields
$f_i(\omega)=-\frac{|\sigma_i \omega|^\alpha}{\alpha !}$ with dispersion parameter $\sigma_i= \|\phi_i\|_{L_\alpha}$. This proves that the wavelet coefficients
of a generalized S$\alpha$S stochastic process follow S$\alpha$S distributions with the spread of the pdf at scale $i$ being determined by the $L_\alpha$ norm of the corresponding
wavelet smoothing kernels. This implies that, for $\alpha<2$, the process is $\ell_\alpha$ compressible in the sense that the essential part of the ``energy content'' is carried by a tiny fraction of wavelet coefficients \cite{Amini2011}.

It should be noted, however, that the quality of the decoupling is strongly dependent upon the spread of the wavelet smoothing kernels $\phi_i$ which should be chosen to be maximally localized for best performance. In the case of the first-order system (cf.\ example in Section \ref{Sec:motiv}), the basis functions for $i$ fixed are not overlapping which implies that the wavelet coefficients within a given scale are independent. This is not so across scale because of the cone-shaped region where the support of the kernels $\phi_{i_1}$ and $\phi_{i_2}$ overlap, which induces dependencies. Incidentally, the inter-scale correlation of wavelet coefficients is often exploited for improving coding performance \cite{Shapiro1993} and signal reconstruction by imposing joint sparsity constraints \cite{Crouse1998}.

%Working with the increment process $s_d(t)$ also has the remarkable feature of completely suppressing long-range dependencies through the B-spline localization process.

%The conclusion of this section is that one can apply standard techniques from system theory (determination of impulse responses) to determine the characteristic form of the whose class of Gaussian and non-Gaussian stationary processes defined by (\ref{eq:ODE}). The functional tools from wavelet theory and exponential B-spline calculus (as we shall see in \cite{Unser2012b})
%%(exponential B-spline calculus) 
%are helpful as well, especially for the handling and characterization of the non-stationary variants of these stochastic processes.

\section{L\'evy processes revisited}
\label{Sec:Levy}
We now illustrate our method by specifying classical L\'evy processes---denoted by $W(t)$---via the solution of the (marginally unstable) stochastic differential equation
\begin{equation}
\label{eq:Levy}
\frac{\dint }{\dint t} W(t) = w(t)
\end{equation}
where the driving term $w$ is one of the independent noise processes defined earlier.
It is important to keep in mind that Eq. (\ref{eq:Levy}), which is the limit of \eqref{eq:AR1w} as $\alpha\to 0$, is only a notation whose correct interpretation is $\langle \Dop W, \varphi \rangle=\langle w, \varphi \rangle$ for all $\varphi \in \mathcal{S}$. 
We shall consider the solution $W(t)$ for all $t\inR$, but we shall impose the boundary condition $W(t_0)=0$ with $t_0=0$ to make our construction compatible with the classical one which is defined for $t\ge0$. 
%When $w$ is Gaussian, the solution of this differential equation turns out to be the classical Wiener process. 

%In writing Eq. (\ref{eq:Levy}), we have intentionally abused the notation in order to be provocative: indeed, when $w$ is Gaussian, the solution of this differential equation turns out to be the classical Wiener process, which is known to be continuous everywhere (almost surely), but nowhere differentiable in the classical sense! Of course, this only makes sense if we interpret (\ref{eq:Levy}) as $\langle \Dop W, \varphi \rangle=\langle w, \varphi \rangle$ for all $\varphi \in \mathcal{S}$.
\subsection{Distributional characterization of L\'evy processes}
The direct application of the operator formalism developed in Section III yields the solution of (\ref{eq:Levy}):
$$
W(t)=\Iop_{0,0} w(t) 
$$
where $\Iop_{0,0}$ is the unique right inverse of $\Dop$ that imposes the required boundary condition at $t=0$. % (via the substraction of an appropriate integration constant).  
The Fourier-based expression of this anti-derivative operator is obtained from the 6th line of Table I by setting $(\omega_0,t_0)=(0,0)$.
By using the properties of the Fourier transform, we obtain the simplified expression
\begin{eqnarray}
\label{eq:antiderivadj}
\Iop_{0,0}\varphi(t) 
%=\int_{\R}  \hat{\varphi}(\omega)   \frac{\displaystyle e^{j \omega t}-1}
%       {j\omega}
%   \frac{\dint{\omega\;\;}}{2 \pi}
  = \left\{\begin{array}{lr}
\int_{0}^t \varphi(\tau) \dint \tau, & t\ge 0 \\
-\int_{t}^0 \varphi(\tau) \dint \tau, & t<0,
\end{array}
\right.
\end{eqnarray}
which allows us to interpret $W(t)$ as the integrated version of $w$ with the proper boundary conditions.
Likewise, we derive the time-domain expression of the adjoint operator
\begin{align}
\label{eq:antiderivast}
\Iop^\ast_{0,0}\varphi(t) 
%&=\int_{\R}  \displaystyle 
%   \frac{\hat{\varphi}(\omega) - 
%   \hat{\varphi}(0)  }
%       {-j\omega}
%    e^{j \omega t}\frac{\dint{\omega\;\;}}{2 \pi}\nonumber \\
    &= \left\{
\begin{array}{lr}
 \int_{t}^\infty \varphi(\tau) \dint \tau, & t\ge0,\\
- \int_{-\infty}^t \varphi(\tau)\dint\tau
, & t< 0. \\
\end{array}
\right.
\end{align}
Next, we invoke Proposition \ref{prop:nonstat}  to obtain the characteristic form of the L\'evy process 
%through the following manipulation 
% We have now identified the unique anti-derivative operator $\Dop_0^{-1}$ which is the left-inverse of $\Dop=\frac{\dint }{\dint t}$ that imposes the required boundary condition at $t=0$ (via the addition of an appropriate integration constant).  This allows us to write the solution of (\ref{eq:Levy}) as $s(t)=\Dop^{-1}_0 w(t)$ and to obtain the characteristic form of L\'evy Process through the following manipulation 
%$\Form_W(\varphi)=\Form_{ w}(\Iop^\ast_{0,0}\varphi)$,
\begin{equation}
\label{eq:charformlevy}
\Form_W(\varphi)=\Form_{ w}(\Iop^\ast_{0,0}\varphi)
\end{equation}
which is %guaranteed to be 
admissible provided that the L\'evy exponent $f$ fullfils the condition in Theorem 3.
%where we are using the defining property $\langle  \Iop_{0,0} w, \varphi \rangle=\langle   w, \Iop_{0,0}^\ast\varphi \rangle$.
%% to move the adjoint anti-derivative operator onto the argument of the characteristic form of the noise. 
%Proposition 2 ensures that $\Iop_{0,0}^\ast\varphi$ is bounded and that it preserves the rapid decay of the test functions $\varphi \in \mathcal{S}$. This allows us to infer that
%the L\'evy characteristic functional is well-defined (continuous and positive-definite) and to invoke the Milos-Bochner Theorem to associate it with a process over $\mathcal{S}'$ (Property 2).

We get the characteristic function of the sample values of the L\'evy process
$W(t_1)=\langle W, \delta(\cdot-t_1) \rangle$ by making the substitution
$\varphi=\omega_1 \delta(\cdot-t_1)$ in (\ref{eq:charformlevy}):
$\Form_W\big(\omega_1 \delta(\cdot-t_1)\big)=\Form_w\big(\omega_1
\Iop_{0,0}^\ast\delta(\cdot-t_1)\big)$ with $t_1>0$. We then use
(\ref{eq:antiderivadj}) to evaluate
$\Iop_{0,0}^\ast\delta(t-t_1)=\One_{[0,t_1)}(t)$. Since the latter indicator
function is equal to one for $t \in [0,t_1)$ and zero elsewhere, it is easy to
evaluate the integral over $t$ in (\ref{eq:gennoise}) with $f(0)=0$, %where \revision{the
%L\'evy exponent $f$} is given by (\ref{eq:levykintchine}), 
which yields
\begin{align*}
\E\{e^{j  \omega_1W(t_1)}\}&=\exp\left(\int_\R f\big(\omega_1 \One_{[0,t_1)}(t)\big) \dint t \right)
\onetwocol{}{\\ &}=e^{t_1 f(\omega_1)}%\exp\left({t_1 f(\omega_1)}\right)
%\\&=&\exp\left(j b''_1 \omega_1t_1  - \frac{b_2 \omega_1^2t_1 }{2}+t_1 \int_{|a|<0} [e^{j a \omega_1} - 1 - j \omega_1 \mathrm{1}_{|a|<1}(a)]  \ V(\dint a)  \right).
\end{align*}
This result is equivalent to the celebrated L\'evy-Khinchine representation of the process \cite{Sato1994}.

\subsection{L\'evy increments vs.\ wavelet coefficients}
A fundamental property of L\'evy processes is that their increments at equally-spaced intervals are i.i.d. \cite{Sato1994}. % with an infinite-divisible probability law. 
To see how this fits into the present framework, we specify the increments on the integer grid as the special case of \eqref{Eq:predict} with $\alpha=0$:
\begin{align*}
u[k]&=\Delta_0W(k):=W(k)-W(k-1)\\
&=\int_{k-1}^k w(t) \dint t=\langle w, \beta^\vee_0(\cdot-k)\rangle
\end{align*}
where $\beta_0(t)=\One_{[0,1)}(t)=\Delta_0 \rho_0(t)$ is the causal B-spline of degree 0 (rectangular function).
We are also introducing some new notation, which is consistent with the definitions given in \cite[Table II]{Unser2012b}, to set the stage for the generalizations to come.
$\Delta_0$ is the finite-difference operator, which is the discrete analog of the derivative operator $\Dop$, while $\rho_0$ (unit step) is the Green function of the derivative operator $\Dop$. The main point of the exercise is to show that determining increments is structurally  equivalent to 
the computation of the wavelet coefficients in Property \ref{Prop:wav1} with the smoothing kernel $\phi_i$ being substituted by $\beta_0^\vee$.
It follows that the characteristic function of $w_d[\cdot]$ is given by 
\begin{align}
\label{eq:charfunlevyinc}
\hat p_{u}(\omega)&=\exp\left(\int_\R f(\omega\beta_0^\vee(t)\big)\dint t\right)=e^{f(\omega)}=\hat p_{\rm id}(\omega)
\end{align}
where the simplification of the integral results from the binary nature of $\beta_{0}$ which is either 1 (on a support of size 1) or zero. This implies that the increments of the L\'evy process are independent (because the B-spline functions $\beta^\vee_0(\cdot-k)$ are non-overlapping) and that their pdf is given by the canonical id distribution of the innovation process $p_{\rm id}(x)$ (cf.\ discussion in Section \ref{Sec:sparse}). 
% The important implication is that the computation of the increments of the samples of a L\'evy process provides a sparse representation of the signal that is fully decoupled and whose statistical distribution is completely known.

The alternative is to expand the L\'evy process in the Haar basis which is ideally matched to it. Indeed, the Haar wavelet at scale $i=1$
(lower-left function in Fig. 2) can be expressed as
\begin{align}
\psi_{\rm Haar}(t/2)&=\beta_0(t)-\beta_0(t-1)=\Delta_0\beta_0=\Dop \beta_{(0,0)}(t)
\end{align}
where $\beta_{(0,0)}=\beta_0 \ast \beta_0$ is the causal B-spline of degree 1 (triangle function). Since $\Dop^\ast=-\Dop$, this confirms that the underlying smoothing kernels are dilated versions of a B-spline of degree $1$. % so that the pdf of wavelet coefficients can be readily derived from the formula in Property . 
Moreover, since the wavelet-domain sampling is critical, there is no overlap of the basis functions within a given scale
which implies that the wavelets coefficients are independent on a scale-by-scale basis (cf. Property \ref{Prop:joint}).
If we now compare the situation with that of the L\'evy increments, we observe that the wavelet analysis involves one more layer of smoothing of the innovation with $\beta_0$ (due to the factorization property of $\beta_{(0,0)}$) which slightly complicates the statistical calculations. 

While the smoothing effect on the innovation is qualitatively the same in both instances, there are fundamental differences, too. In the wavelet case, the underlying discrete transform is orthogonal, but the coefficients are not fully decoupled because of the inter-scale dependencies which are unavoidable, as explained in Section \ref{Sec:wav}.
By contrast, the decoupling of the L\'evy increments is perfect, but the underlying discrete transform (finite difference transform) is non-orthogonal. In our companion paper, we shall see how this latter strategy is extendable to the much broader family of  sparse processes via the definition of the {\em generalized
increment process}.
\subsection{Examples of L\'evy processes}
\begin{figure*}
\centering
  \includegraphics[scale=0.7]{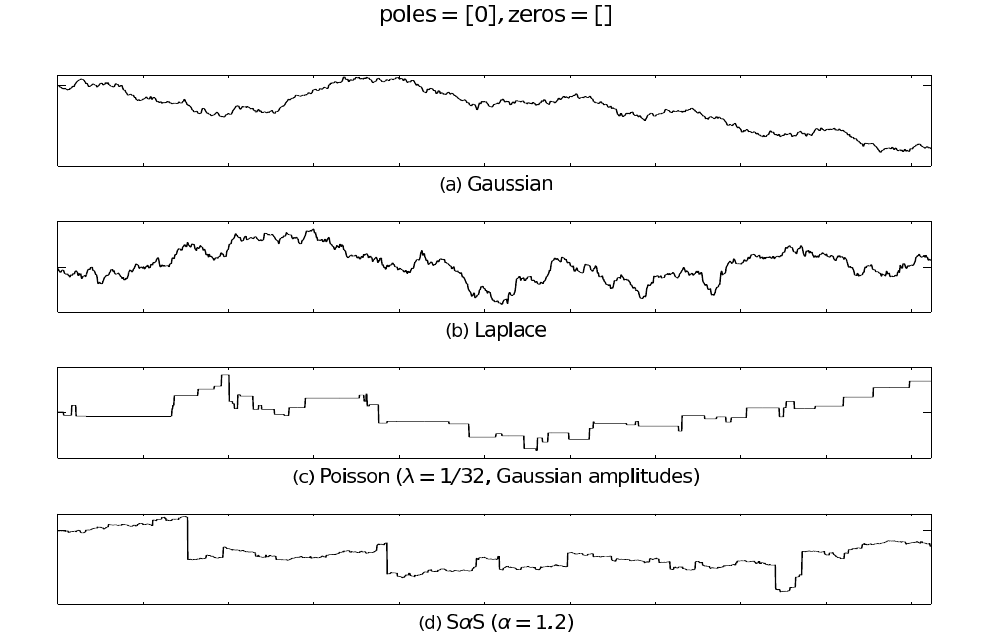}
  \caption{\label{fig:Levy}
Examples of L\'evy motions $W(t)$ with increasing degrees of sparsity. (a) Brownian motion with L\'evy triplet $(0,1,0)$. (b) L\'evy-Laplace motion with $\big(0,0,\frac{e^{-|a|}}{|a|}\big)$. (c) Compound Poisson process with $\big(0,0,\lambda \frac{1}{\sqrt{2 \pi}} e^{-a^2/2}\big)$ with $\lambda=\frac{1}{32}$. (d) Symmetric L\'evy flight with $\big(0,0,1/|a|^{\alpha+1}\big)$ and $\alpha=1.2$. }
\end{figure*}

Realizations of four different L\'evy processes are shown in Fig. 1 together with their L\'evy triplets $\big(b_1,b_2,v(a)\big)$.
The first signal is a Brownian motion (a.k.a. Wiener process) that is obtained by integration of a white Gaussian noise. This classical process is known to be nowhere differentiable in the classical sense, despite the fact that it is continuous everywhere (almost surely) as all the members of the L\'evy family.
While the sampled version of $\Delta_0 W$ is i.i.d. in all cases, it does not yield a sparse representation in this first instance because the underlying distribution remains Gaussian. The second process, which may be termed L\'evy-Laplace motion, is specified by the L\'evy density $v(a)=e^{-|a|}/{|a|}$ which is not in $L_1$. By taking the inverse Fourier transform of (\ref{eq:charfunlevyinc}), we can show that its increment process has a Laplace distribution \cite{Unser2011}; note that this type of generalized Gaussian model is often used to justify sparsity-promoting signal processing techniques based on $\ell_1$ minimization \cite{Bouman1993,Seeger2008,Babacan2010}.
The third piecewise-constant signal is a compound Poisson process. It is intrinsically sparse since a good proportion of its increments is zero by construction (with probability $e^{-\lambda}$). The fourth example is an alpha-stable L\'evy motion (a.k.a. L\'evy flight) with $\alpha=1.2$. Here, the distribution of $\Delta_0 W$ is heavy-tailed (S$\alpha$S) with unbounded moments for $p>\alpha$. Although this may not be obvious from the picture, this is the sparsest process of the lot because it is $\ell_\alpha$-compressible in the strongest sense \cite{Amini2011}. Specifically, we can compress the sequence such as to preserve any prescribed portion $r<1$ of its average $\ell_\alpha$ energy by retaining an arbitrarily small fraction of samples as the length of the signal goes to infinity.
\subsection{Link with conventional stochastic calculus}
Thanks to (\ref{eq:Levy}), we can view a white noise $w=\dot W$ as the weak derivative of some classical L\'evy processes $W(t)$ which is well-defined pointwise (almost everywhere).  This provides us with further insights on the range of admissible white noise processes of Section II.C which constitute the driving terms of the general stochastic differential equation (\ref{eq:operatoreq}). This fundamental observation also makes the connection with stochastic calculus\footnote{The It\^o integral of conventional stochastic calculus is based on Brownian motion, but the concept can also be generalized to L\'evy driving terms using the more advanced theory of semimartingales\cite{Protter2004}.} \cite{Protter2004,Brockwell2001}, which avoids the notion of white noise by relying on the use of stochastic integrals of the form 
$$
s(t)=\int_\R \rho(t,t') \dint W(t')
$$  where $W$ is a random (signed) measure associated to some canonical Brownian motion (or, by extension, a L\'evy process) and where $\rho(t,t')$ is an integration kernel that formally corresponds to our inverse operator $\Lop^{-1}$. 

\section{Conclusion}
We have set the foundations of a unifying framework that gives access to the broadest possible class of continuous-time stochastic processes  specifiable by linear, shift-invariant equations, which is beneficial for signal processing purposes. We have  shown that these processes admit a concise representation in a wavelet-like basis. We have applied our framework to the description of the classical L\'evy processes, which, in our view, provide the simplest and most basic examples of sparse processes, despite the fact that they are non-stationary. % (single pole at the origin). 
We  have also hinted at the link between L\'evy increments and splines, which is the theme that we shall develop in full generality next \cite{Unser2012b}.

%Interestingly, the higher-order extension of L\'evy processes suggested by our formulation (differential system with a pole of multiplicity $N$ at the origin) are sparse processes that are best best represented conventional wavelets \footnote{A wavelet that has $N$ vanishing moments can always be written as $\psi=\Dop^N \phi$ where the operator $\Dop^N$ is scale-invariant.} 
%
% despite the fact that they are non-stationary (due to their pole at the origin). Interestingly, the processes that are best represented in terms of conventional wavelets\footnote{A wavelet that has $N$ vanishing moments can always be written as $\psi=\Dop^N \phi$ where the operator $\Dop^N$ is scale-invariant.} are the special instance of our model with a pole of multiplicity $N$ at the origin. The processes, which are non-stationary, are the only members of our new extended family which are (second-order) self-similar with fractal properties.
%
%
We have demonstrated that the proposed class of stochastic models and the corresponding mathematical machinery (Fourier analysis, characteristic functional, and B-spline calculus) lends itself well to the derivation of transform-domain statistics.  The formulation suggests
 a variety of new processes whose properties are compatible with the currently-dominant paradigm in the field which is focused on the notion of sparsity.
In that respect, the sparse processes that are best matched to conventional wavelets\footnote{A wavelet with $N$ vanishing moments can always be rewritten as $\psi=\Dop^N \phi$ with $\phi\in L_2(\R)$ where the operator $\Lop=\Dop^N$ is scale-invariant.} 
are those generated by $N$-fold integration (with proper boundary conditions) %(i.e., $\Iop_0^N w$) 
of a non-gaussian innovation.
These processes, which are the solution of an unstable SDE (pole of multiplicity $N$ at the origin), are intrinsically self-similar (fractal) and non-stationary.
%: they are non-classical entities that would have been much harder to pin down using conventional stochastic calculus. 
Last but not least, the formulation is backward compatible with the classical theory of Gaussian stationary processes.

%is compatible with the notions of linearity and (quasi) shift-invariance which are so central to signal processing and system theory.
\appendices

\section*{Appendix I: Positive-definite functionals}
We start by recalling the fundamental notion of positive-definiteness for univariate functions \cite{Stewart1976}. 
\begin{definition} A complex-valued function $f$ of the real variable $\omega$ is said to be {\em positive-definite} iff.
$$
\sum_{m=1}^N \sum_{n=1}^N f(\omega_m-\omega_n) \xi_m\overline{\xi}_n \ge 0
$$
for every possible choice of $\omega_1,\dots,\omega_N \inR$, $\xi_1,\dots,\xi_N \in \C$ and $N \in \Z_+$. 
\end{definition}
This is equivalent to the requirement that the $N\times N$ matrix ${\bf F}$ whose elements are given by
$[{\bf F}]_{mn}=f(\omega_m-\omega_n)$ is positive semi-definite (that is, non-negative definite) for all $N$, no matter how the $\omega_n$'s are chosen.

Bochner's theorem states that a bounded, continuous function $f$ is positive-definite if and only if it is the Fourier transform of a positive and finite Borel measure $\mu$:
$$f(\omega)=\int_{\R} e^{j \omega x} \mu(\dint x).$$
In particular, Bochner's theorem implies that $f$ is a valid characteristic function---that is, $f(\omega)=\E\{ e^{j  \omega X}\}=\int_{\R} e^{j \omega x}  \mu(\dint x)$ where $X$ is a random variable with probability measure $\Meas_X=\mu$---iff. $f$ is continuous, positive-definite and $f(0)=1$. Note that the above results and formulas also generalize to the multivariate setting.

These concepts carry over as well to functionals on some abstract nuclear space $\Spc X$, the prime example being Schwartz's class $\mathcal{S}$ of smooth and rapidly-decreasing test functions\cite{Gelfand-Villenkin1964}.

\begin{definition} A complex-valued functional $L(\varphi)$ defined over the function space $\Spc X$ is said to be {\em positive-definite} iff.
$$
\sum_{m=1}^N \sum_{n=1}^N L(\varphi_m-\varphi_n) \xi_m\xi^\ast_n \ge 0
$$
for every possible choice of $\varphi_1,\dots,\varphi_N \in \Spc X$, $\xi_1,\dots,\xi_N \in \C$ and $N \in \N^+$. 
\end{definition}

\begin{theorem} [Minlos-Bochner]
\label{Th:Minlos}
Given a functional $\Form_s(\varphi)$ on a nuclear space $\Spc X$ that is continuous, positive-definite and such that $\Form_s(0)=1$, there exists a unique probability measure $\Meas_s$ on the dual space $\Spc X'$ such that
$$
\Form_s(\varphi)=\E\{e^{j \langle s, \varphi \rangle}\}=\int_{\Spc X'} e^{j \langle s, \varphi \rangle} \dint \Meas_s(s),
$$
where $\langle s, \varphi \rangle$ is the dual pairing map. One further has the guarantee that all finite dimensional probabilities measures   derived from $\Form_s(\varphi)$  by setting $\varphi= \omega_1 \varphi_1+ \cdots + \omega_N \varphi_N$ are mutually compatible.
\end{theorem}

The characteristic form therefore uniquely specifies the generalized stochastic process $s=s(\varphi)$ (via the infinite-dimensional probability measure $\Meas_s$) in essentially the same way as the characteristic function fully determines the probability measure of a scalar or multivariate random variable.
%
%The generalized white noise functionals (\ref{eq:gennoise}) where $f$ is given by (\ref{eq:levykintchine}) satisfy the required conditions for $\varphi \in \mathcal{S}$ and therefore specify bona fide generalized stochastic processes whose elements are in $\mathcal{S}'$.
%
%In order to construct derived processes resulting from the applycation of a linear transformation $T$, we need to make sure that the action of the adjoint operator $\Top^\ast$ on $\varphi\in\mathcal{S}$ preserves the continuity of the functional $Z_s(\varphi)=Z_w(\Top^\ast\varphi)$. This will obviously be the case whenever $\Top^\ast$ is $\mathcal{S}$-continuous.
%Otherwise, one needs to show that $\Top^\ast$ satisfies some $L_p$-boundedness condition, as specified in Theorem 3.

\section*{Appendix II: Proof of Theorem 3}
 1) \quad
% For the First the linearity of $T\Phi$, i.e.,
%\begin{equation}\label{linearity.n}
%T^*_n\Phi(\alpha f+\beta g)=\alpha T^*_n\Phi(f)+\beta T^*_n
%\Phi(g)
%\end{equation}
%for any $f, g\in {\mathcal D}$ and $\alpha, \beta\in \RR$. By
%\eqref{tn.def} and the linearity of the random process $\Phi$ and
%the linear operator $T_n$, we have that
%\begin{eqnarray*}
%T^*_n \Phi(\alpha f+\beta g)\nonumber
%&=& \Phi( T_n(\alpha f+\beta g)) \quad {\rm by \ \eqref{tn.def}}\nonumber\\
%&=&
%\Phi(\alpha T_n f+\beta T_n g) \quad {\rm by \ the\ linearity\ of\ the\ operator} \ T_n\nonumber\\
%&=& \alpha \Phi(T_n f)+\beta \Phi(T_n g)\nonumber\\
% & & \qquad \quad {\rm by\ the\ linearity\ of\ the\ generalized \ random \ process } \ \Phi\nonumber\\
%& = & \alpha T^*_n \Phi(f)+\beta T^*_n\Phi(g) \quad {\rm by\
%\eqref{tn.def}}
%\end{eqnarray*}
%Hence  \eqref{linearity.n}  follows.
As $w$ is a generalized random process, $\Form_w$ is a
continuous functional on ${\mathcal S}$. This, together with the
assumption that $\Top$ is a continuous operator on ${\mathcal S}$,
implies that the composed functional $\Form_s(\varphi):=\Form_w(\Top\varphi)$  is continuous on ${\mathcal S}$.

Given  the functions $\varphi_1, \ldots, \varphi_N$ in ${\mathcal S}$
and some complex coefficients $\xi_1, \ldots, \xi_N$,
\begin{align*}
\sum_{1\le m, n\le N} &\Form_s(\varphi_m-\varphi_{n})
 \xi_m \overline{\xi_{n}}\\
= & \sum_{1\le m,n\le N} \Form_w\big(\Top(\varphi_m-\varphi_{n})\big)
 \xi_m \overline{\xi_{n}} \\
 = &  \sum_{1\le m,n\le N} \Form_w(\Top\varphi_m-\Top\varphi_n)
 \xi_m \overline{\xi_{n}} \tag{\rm by \ the \ linearity\ of \ the \ operator\ $T$}\nonumber\\
 \ge &  0 \hfill \tag{\rm by\ the\ positivity\ of \ ${\mathcal
Z}_w$}.
\end{align*}
This proves the positive-definiteness of the functional $\Form_s$ on ${\mathcal S}$.

 Clearly, $\Form_s(0)={\mathcal
Z}_w(\Top 0)=\Form_w(0)=1$.

2)\quad By the \revision{continuity of the operator $\Top$ from $\Spc S$ into
$L_p$}, $\Top\varphi\in L_p$
for all $\varphi \in {\mathcal S}$. This together with the
assumption  $|f(u)|\le C |u|^p$ implies that $\Form_s(\varphi)=\exp(\int_{\Bbb R} f(\Top\varphi(t))\dint t)$ is
well-defined for all $\varphi\in {\mathcal S}$. By the linear
property of the operator $\Top$ and $f(0)=0$, we obtain that
$\Form_s(0)=1$. The positive-definiteness of the functional
$\Form_s$ is established by an argument similar to the one used above.
Finally we prove the continuity of the functional $\Form_s$
on ${\mathcal S}$: Let $\{\varphi_n\}_{n=1}^\infty$ be a
convergent sequence in ${\mathcal S}$ and denote its limit in
${\mathcal S}$ by $\varphi$. Then by the assumption on the linear
operator $\Top$, $\Top \varphi_n$ converges to $\Top\varphi$ in $L_p$; that
is,
\begin{equation}\label{fnlimit.eq}
\lim_{n\to \infty} \|\Top\varphi_n-\Top \varphi\|_p=0.
\end{equation}
Next, we observe that
\begin{align*}
|f(u)-f(v)|  = & \Big|\int_{v}^u f'(t)\dint t\Big|
%\quad {\rm (by \
%the\  fundamental\ theorem)}
\\
  \le &  C \Big|\int_v^u t^{p-1} \dint t\Big| \tag {\rm by \ the \ assumption on \ $f$}\\
   \le &  C \max(|u|^{p-1}, |v|^{p-1}) |u-v| \\
 \le &  C (|v|^{p-1}+|u-v|^{p-1})|u-v|.  \tag {\rm by \ the \
triangle\ inequality}
\end{align*}
We then have
\begin{align*}
 & \Big|\int_{\Bbb R} f(\Top\varphi_n(t)) \dint t-\int_{\Bbb R} f(\Top
\varphi(t)) \dint t\Big|\\
 \le & C  \int_{\Bbb R} |\Top\varphi(t)|^{p-1}
|\Top\varphi_n(t)-\Top\varphi(t)| + |\Top\varphi_n(t)-\Top\varphi(t)|^p \dint t\\
 \le & C \Big(\|\Top\varphi\|_p^{p-1} \|\Top\varphi_n-\Top\varphi\|_p+
\|\Top\varphi_n-\Top\varphi\|_p^p\Big) \tag{\rm by \ H\"older's \ inequality}\\
 \to & 0 \ {\rm as} \ n\to \infty, \tag {\rm by \
\eqref{fnlimit.eq}}
\end{align*}
which proves the continuity of the functional $\Form_s$ on
${\mathcal S}$.
\endproof

%It\^o's regularization theorem [\cite[Theorem 2.3.2]{Ito1984} allows one to extend the range of applicability of the above existence theorem by considering processes on $\mathcal{S}$ induced by linear transformation of regular ones (as defined above), provided that the dual operator is a continuous map on $\mathcal{S}$ and not necessarily a mapping from $\mathcal{S}$ into itself.

\section*{Acknowledgements}
\revision{ The research was partially supported by the Swiss National Science
Foundation under Grant 200020-109415, the European Commission under Grant
ERC-2010-AdG 267439-FUN-SP, and the National Science Foundation under Grant
DMS 1109063.}
The authors are thankful to Prof. Victor Panaretos (EPFL chair of Mathematical Statistics) and Prof. Robert Dalang (EPFL Chair of Probabilities) for helpful discussions.

%\vspace*{-2ex}
%\vspace{-0.2cm}
\bibliographystyle{IEEEtran}
%\bibliography{/Users/munser/Bibliography/Bibtex_files/monogenic,/Users/munser/Bibliography/Bibtex_files/BIG}
\bibliography{/Users/munser/Bibliography/Bibtex_files/Unser}

\end{document}